\documentclass[12pt,amsmath,amssymb,aip,jcp,preprint,floatfix]{revtex4-1}
\usepackage{graphicx}
\usepackage{color}
\usepackage[version=3]{mhchem}
\usepackage{dcolumn}
\usepackage{multirow, makecell}
\usepackage[normalem]{ulem}
\usepackage{cmbright}
\usepackage[dvipsnames]{xcolor}
\usepackage{array}
\usepackage{siunitx}
\usepackage{textgreek}
\usepackage{csquotes}
\usepackage{longtable}
\usepackage{comment}
\usepackage{graphics}

\usepackage{rotating}      
\usepackage{tabularray}    
\usepackage{graphicx}      
\usepackage{xcolor}        
\usepackage{threeparttable}   
\usepackage{ragged2e} 

\usepackage{ulem}

\begin{document}

\newcommand\bfec[1]{\textcolor{orange}{\textit{ XXX #1 XXX }}}
\newcommand\jj[1]{\textcolor{cyan}{ JJ: #1 }}
\newcommand\fede[1]{\textcolor{magenta}{#1}}

\newcommand{\nhl}[1]{{\color{blue} NHL: #1 }}
\newcommand\notetoauthor[1]{{\color{red} NB: #1 }}

\setlength{\tabcolsep}{12pt}

\DeclareFontFamily{OT1}{cmbr}{\hyphenchar\font45 }
\DeclareFontShape{OT1}{cmbr}{m}{n}{%
  <-9>cmbr8
  <9-10>cmbr9
  <10-17>cmbr10
  <17->cmbr17
}{}
\DeclareFontShape{OT1}{cmbr}{m}{sl}{%
  <-9>cmbrsl8
  <9-10>cmbrsl9
  <10-17>cmbrsl10
  <17->cmbrsl17
}{}
\DeclareFontShape{OT1}{cmbr}{m}{it}{%
  <->ssub*cmbr/m/sl
}{}
\DeclareFontShape{OT1}{cmbr}{b}{n}{%
  <->ssub*cmbr/bx/n
}{}
\DeclareFontShape{OT1}{cmbr}{bx}{n}{%
  <->cmbrbx10
}{}

\renewcommand{\rmdefault}{cmbr}
\renewcommand{\sfdefault}{cmbr}

\renewcommand{\thesection}{\arabic{section}}
\renewcommand{\thesubsection}{\thesection.\arabic{subsection}}
\renewcommand{\thesubsubsection}{\thesubsection.\arabic{subsubsection}}

\makeatletter
\renewcommand{\p@subsection}{}
\renewcommand{\p@subsubsection}{}
\makeatother

\title{Perspective on a challenge: predicting the photochemistry of cyclobutanone}


\author{Ji\v{r}\'{i} Jano\v{s}}%
\affiliation{School of Chemistry, University of Bristol, Bristol BS8 1TS, United Kingdom}%
\affiliation{Department of Physical Chemistry, University of Chemistry and Technology, Prague, Technick\'{a} 5, 16628 Prague, Czech Republic}

\author{Nanna Holmgaard List}
\affiliation{School of Chemistry, University of Birmingham, Birmingham B15 2TT, United Kingdom}
\affiliation{Department of Chemistry, KTH Royal Institute of Technology, SE-10044 Stockholm, Sweden}

\author{Andrew J. Orr-Ewing}%
\affiliation{School of Chemistry, University of Bristol, Bristol BS8 1TS, United Kingdom}%

\author{Jiří Suchan}
\affiliation{Institute for Advanced Computational Science, Stony Brook University, Stony Brook, New York 11794, United States}

\author{Mario Barbatti}
\affiliation{Aix Marseille University, CNRS, ICR, 13397 Marseille, France}
\affiliation{Institut Universitaire de France, 75231 Paris, France}

\author{Olivia Bennett}
\affiliation{Department of Chemistry, University College London, 20 Gordon St., WC1H 0AJ London, United Kingdom}

\author{Marcus Brady}
\affiliation{Department of Chemistry, University College London, 20 Gordon St., WC1H 0AJ London, United Kingdom}

\author{Javier Carmona-García}
\affiliation{School of Chemistry, University of Bristol, Bristol BS8 1TS, United Kingdom}%

 \author{Rachel Crespo-Otero}
\affiliation{Department of Chemistry, University College London, 20 Gordon St., WC1H 0AJ London, United Kingdom}

\author{Julien Eng}
\affiliation{Chemistry, School of Natural and Environmental Sciences, Newcastle University, Newcastle Upon Tyne NE1 7RU, United Kingdom}

\author{O. Jonathan Fajen}
\affiliation{Department of Chemistry and The PULSE Institute, Stanford University, Stanford, California 94305, United States}
\affiliation{SLAC National Accelerator Laboratory, 2575 Sand Hill Road, Menlo Park, California 94025, United States}

\author{Marco Garavelli}
\affiliation{Dipartimento di Chimica industriale ``Toso Montanari'', Università di Bologna, Viale del Risorgimento 4, 40136 Bologna, Italy}

\author{Sandra Gómez}
\affiliation{Departamento de Química, Módulo 13, Universidad Autónoma de Madrid, 28049 Madrid, Spain}

\author{Alice E. Green}
\affiliation{EaStCHEM School of Chemistry, University of Edinburgh, Edinburgh EH9 3FJ, United Kingdom}

\author{Federico J. Hernández}
\affiliation{Department of Chemistry, Queen Mary University of London, Mile End Road, London E1 4NS, United Kingdom}%

\author{Daniel Hollas}
\affiliation{School of Chemistry, University of Bristol, Bristol BS8 1TS, United Kingdom}%

\author{Lewis Hutton}
\affiliation{Physical and Theoretical Chemistry Laboratory, Department of Chemistry, University of Oxford, Oxford OX1 3QZ, United Kingdom}

\author{Lea M. Ibele}
\affiliation{Aix Marseille University, CNRS, ICR, 13397 Marseille, France}

\author{Adam Kirrander}
\affiliation{Physical and Theoretical Chemistry Laboratory, Department of Chemistry, University of Oxford, Oxford OX1 3QZ, United Kingdom}
 
\author{Zhenggang Lan}
\affiliation{SCNU Environmental Research Institute, Guangdong Provincial Key Laboratory of Chemical Pollution and Environmental Safety \& MOE Key Laboratory of Environmental Theoretical Chemistry, School of Environment, South China Normal University, Guangzhou 510006, China}

\author{Yorick Lassmann}
\affiliation{School of Chemistry, University of Bristol, Bristol BS8 1TS, United Kingdom}%

\author{Joseph E. Lawrence}
\affiliation{Simons Center for Computational Physical Chemistry, New York University, New York, NY 10003, United States}
\affiliation{Department of Chemistry, New York University, New York, NY 10003, United States}

\author{Benjamin G. Levine}
\affiliation{Institute for Advanced Computational Science, Stony Brook University, Stony Brook, New York 11794, United States}
\affiliation{Department of Chemistry, Stony Brook University, Stony Brook, New York 11794, United States}

\author{Dmitry V. Makhov}
\affiliation{School of Chemistry, University of Leeds, Leeds, LS2 9JT, United Kingdom}

\author{Jonathan R. Mannouch}
\affiliation{Hamburg Center for Ultrafast Imaging, Universität Hamburg and Max Planck Institute for the Structure and Dynamics of Matter, Luruper Chaussee 149, 22761 Hamburg, Germany}

\author{Xincheng Miao}
\affiliation{Institut für Physikalische und Theoretische Chemie, Julius-Maximilians-Universität Würzburg, Emil-Fischer-Straße 42, 97074 Würzburg, Germany}

\author{Roland Mitrić}
\affiliation{Institut für Physikalische und Theoretische Chemie, Julius-Maximilians-Universität Würzburg, Emil-Fischer-Straße 42, 97074 Würzburg, Germany}


\author{Shane M. Parker}
\affiliation{Department of Chemistry, Case Western Reserve University, Cleveland, OH, United States}

\author{Thomas J. Penfold}
\affiliation{Chemistry, School of Natural and Environmental Sciences, Newcastle University, Newcastle Upon Tyne NE1 7RU, United Kingdom}

\author{Jiawei Peng}
\affiliation{SCNU Environmental Research Institute, Guangdong Provincial Key Laboratory of Chemical Pollution and Environmental Safety \& MOE Key Laboratory of Environmental Theoretical Chemistry, School of Environment, South China Normal University, Guangzhou 510006, China}

\author{Jeremy O. Richardson}
\affiliation{Department of Chemistry and Applied Biosciences, ETH Zurich, 8093 Zurich, Switzerland}

\author{Dmitrii Shalashilin}
\affiliation{School of Chemistry, University of Leeds, Leeds, LS2 9JT, United Kingdom}

\author{Petr Slavíček}
\affiliation{Department of Physical Chemistry, University of Chemistry and Technology, Prague, Technick\'{a} 5, 16628 Prague, Czech Republic}

\author{K. Eryn Spinlove}
\affiliation{Department of Chemistry, University College London, 20 Gordon St., WC1H 0AJ London, United Kingdom}

\author{Patricia Vindel-Zandbergen}
\affiliation{Department of Chemistry, New York University, New York, NY 10003, United States}

\author{Federica Agostini}
\affiliation{Universit\'e Paris-Saclay, CNRS, Institut de Chimie Physique UMR8000, 91405, Orsay, France}
\affiliation{Sorbonne Universit\'e, CNRS, LCT UMR 7616, Paris 75005, France}

\author{Sara Bonella}
\affiliation{Centre Européen de Calcul Atomique et Moléculaire (CECAM), École Polytechnique Fédérale de Lausanne, 1015 Lausanne, Switzerland}

\author{Todd J. Martínez}
\affiliation{Department of Chemistry and The PULSE Institute, Stanford University, Stanford, California 94305, United States}
\affiliation{SLAC National Accelerator Laboratory, 2575 Sand Hill Road, Menlo Park, California 94025, United States}

\author{Graham A. Worth}
\affiliation{Department of Chemistry, University College London, 20 Gordon St., WC1H 0AJ London, United Kingdom}

\author{Basile F. E. Curchod}
\email{basile.curchod@bristol.ac.uk}
\affiliation{School of Chemistry, University of Bristol, Bristol BS8 1TS, United Kingdom}%

\date{\today}%

\begin{abstract}
This Perspective is part of a Special Topic that explored the maturity of nonadiabatic molecular dynamics for predicting photochemical processes. In 2023, a prediction challenge was issued to the community of computational photochemists to simulate the photochemistry of cyclobutanone, photoexcited at 200 nm, and the resulting time-resolved MeV-UED signal. The challenge attracted 15 theoretical predictions from more than 70 researchers, employing a wide range of strategies for electronic structure and nonadiabatic molecular dynamics to predict the time-resolved MeV-UED signal \textit{before the experiment had been conducted} at SLAC (Stanford, USA). The MeV-UED instrument at Shanghai Jiao Tong University was also used to provide a second independent time-resolved MeV-UED signal for the photochemistry of cyclobutanone.

This Perspective discusses the various approaches and strategies used by the participants to predict the photochemistry of cyclobutanone. This work also summarizes the strengths and weaknesses of various methods used for photoexcitation, electronic structure, nonadiabatic dynamics, and calculation of observables, as agreed by the participants during a CECAM workshop dedicated to the results of the challenge and organized in Lausanne in April 2025. This Perspective also collects all the predicted time-resolved MeV-UED signals into a single figure, together with the experimental signal. 
This challenge (i) demonstrated the qualitative predictive power of nonadiabatic molecular dynamics and (ii) underscored the impact of electronic-structure theory on the outcome of the excited-state dynamics and the need for its careful benchmarking. This effort allowed the community to share practical strategies to perform nonadiabatic dynamics (discussed in the present Perspective) and constitutes a 'calibration' exercise for computational photochemistry. 
\end{abstract}

\maketitle

\tableofcontents
\clearpage

\section{Introduction}
\label{sec:introduction}

Developing methods to simulate the dynamics of electronically excited molecules beyond the Born--Oppenheimer approximation, coined \textit{nonadiabatic molecular dynamics}, has been a very active field of research over the past decades with numerous applications, e.g., in spectroscopy,\cite{liu2025reviewtoddelectrocyclic} photochemistry,\cite{levine2007isomerization,schuurman2018conical} or atmospheric chemistry.\cite{curchod2024atmo} Nonadiabatic molecular dynamics relies not only on strategies that can capture the subtle coupling between electronic and nuclear motions in regions of strong nonadiabaticity, but also on the possibility to approximate some of the nuclear quantum effects unraveled by the photoexcitation process. Recent applications of nonadiabatic molecular dynamics to realistic simulations of molecules also highlighted the key importance of the underlying electronic-structure methods,\cite{Janos2023,papineau2024elecstrucnamd,jirasensitivityNAMD2024} which provide electronic energies, nuclear forces, nonadiabatic coupling matrix elements, and other electronic quantities required for the nuclear propagation. Photophysical processes, where the nuclear configuration space visited by the photoexcited molecule does not include bond breakage or formation, can often be well described by simple, single-reference quantum-chemical methods. Conversely, photochemical processes, which span a much larger region of the nuclear configuration space, typically require electronic-structure methods including both static and dynamic correlation. The methods picked for the nonadiabatic dynamics and electronic structure are, however, not the only important ingredients for a successful simulation of excited-state processes: the description of the photoexcitation process and the calculation of experimental observables of interest are key components for a direct comparison with experiments. Hence, performing nonadiabatic molecular dynamics requires control over a vast set of parameters for different highly entangled components: photoexcitation, electronic structure, nonadiabatic dynamics, and calculation of observables.\cite{cigrang2025roadmapNAMD,bestpractices2026}

\begin{figure}[ht]
\centering
\includegraphics[width=0.75\textwidth]{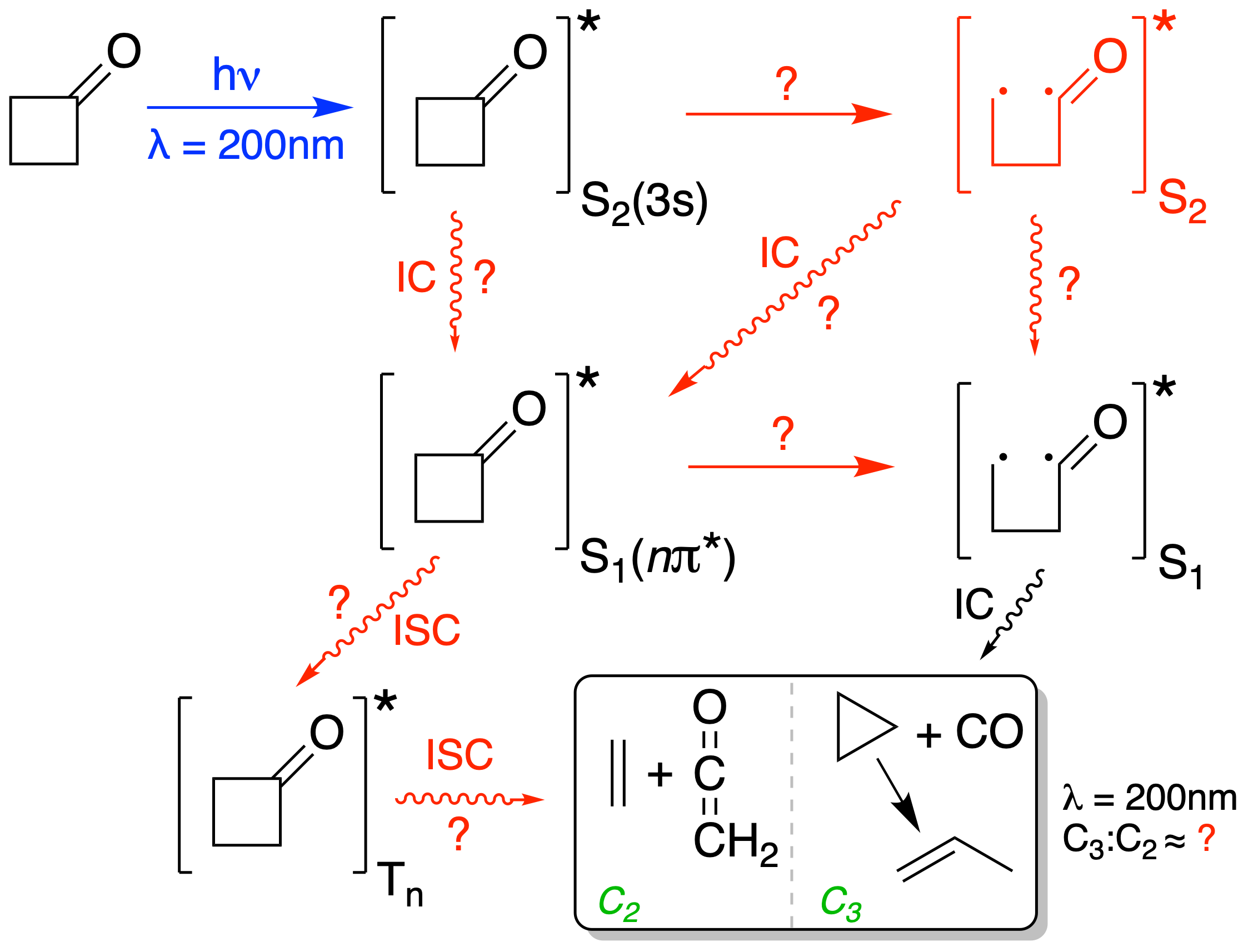}
\caption{Schematic representation of the potential photochemistry of cyclobutanone triggered by photoexcitation at 200 nm (adapted from Ref.~\citenum{curchod2024cyclo}). Red symbols indicate the pathways discussed and questions addressed as part of the Prediction Challenge. The C$_\text{2}$ and C$_\text{3}$ photoproducts are highlighted. We note that ethenone (C$_\text{2}$ product) can possibly undergo further reaction to produce \ce{CO} and \ce{CH2}.}
\label{fig:schemecyclo}
\end{figure}

The challenge of combining the various ingredients necessary for nonadiabatic molecular dynamics has often constrained excited-state simulations to act as an interpretative tool for (time-resolved) spectroscopy and photochemistry. This Special Topic asked the question: \textit{Is the field of nonadiabatic molecular dynamics mature enough to be predictive?} This question emerged during a discussion session of the CECAM workshop entitled 'Triggering out-of-equilibrium dynamics in molecular systems', which took place in Lausanne (Switzerland) between the 28th and the 31st of March 2023. Motivated to find an answer to this question, a subgroup of participants in this workshop developed the idea of this prediction challenge. Discussion with scientists at the SLAC Megaelectronvolt Ultrafast Electron Diffraction facility (Stanford, USA) led to the identification of an experiment scheduled for the end of January 2024: a time-resolved mega-electronvolt ultrafast electron diffraction (MeV-UED) study of the photochemistry of cyclobutanone, photoexcited at 200 nm. Such an experiment was considered a good fit for a challenge: (i) the knowledge about the photochemistry of cyclobutanone excited at 200 nm in the gas phase is limited (Fig.~\ref{fig:schemecyclo}), yet it is expected that photoproducts (resulting from photodissociation pathways) are formed, (ii) intersystem crossing may play a role in the excited-state dynamics, (iii) time-resolved MeV-UED offers both spatial and temporal resolution at the atomic scale, i.e., molecular movies -- a great way to monitor the formation of photoproducts.

The Journal of Chemical Physics (JCP) agreed to host the articles resulting from this challenge as part of this Special Topic and sent invitation emails to members of the community at the end of July 2023. The challenge consisted of predicting (i) the photochemistry of gas-phase cyclobutanone, photoexcited by a 200 nm laser pulse, and (ii) the time-resolved mega-electronvolt ultrafast electron diffraction (MeV-UED) signal that would be recorded at SLAC. Point (ii) was a central requirement of this call, ensuring a one-to-one comparison between theory and experiment. All participants were informed of the earlier experimental and theoretical literature on this molecule.\cite{kao2020effects,diau2001femtochemistry,xia2015excited,liu2016new} The deadline was set to the 15th of January 2024, subsequently extended to the 15th of February 2024. 15 articles resulted from this call, gathering more than 70 researchers to address this challenge. The range of strategies employed for the electronic structure and the nonadiabatic dynamics was broad and will be dissected in Sec.~\ref{sec:challenge}. 

The time-resolved MeV-UED experimental article published in this Special Topic proposed a detailed analysis of the recorded signals without using the results of the nonadiabatic dynamics simulations part of the challenge.
This experiment (Exp1 in the following), was conducted at the SLAC Megaelectronvolt Ultrafast Electron Diffraction facility and sampled the photodynamics of cyclobutanone up to 2 ps following photoexcitation (200 nm, 130 fs FWHM).\cite{green2025cycloUEDExp1} Exp1 interpreted their observations as showing that cyclobutanone is initially photoexcited to a $3s$-Rydberg state, which decays within (0.29 $\pm$ 0.2) ps via a rapid ring-opening (Norrish Type-I) reaction. The ring-opening product fragments within the experimental time window to form mainly ketene and ethylene (so-called C$_\text{2}$ product channel) or propene/cyclopropane and CO (so-called C$_\text{3}$ product channel). 

It is exciting that the Prediction Challenge also stimulated a second time-resolved MeV-UED experiment on cyclobutanone (Exp2), performed with the MeV UED instrument at Shanghai Jiao Tong University and focusing on the first 1.2 ps of dynamics following photoexcitation of cyclobutanone (199.5 nm, 120 fs FWHM).\cite{wang2025cycloUEDExp2} Exp2 observed an excited-state lifetime for the $3s$-Rydberg state of about (0.23 $\pm$ 0.04) ps, followed by dissociation. The main photoproducts are found to be cyclopropane and CO, with a branching ratio between the C$_\text{3}$ and C$_\text{2}$ channels of approximately 5:3. The two photoproducts account for $\sim$ 80\% of the photoproducts observed within the experimental time window. Additional discussions on these two experiments are provided in Sec.~\ref{sec:obs}.

\begin{figure}[ht]
\centering
\includegraphics[width=1.0\textwidth]{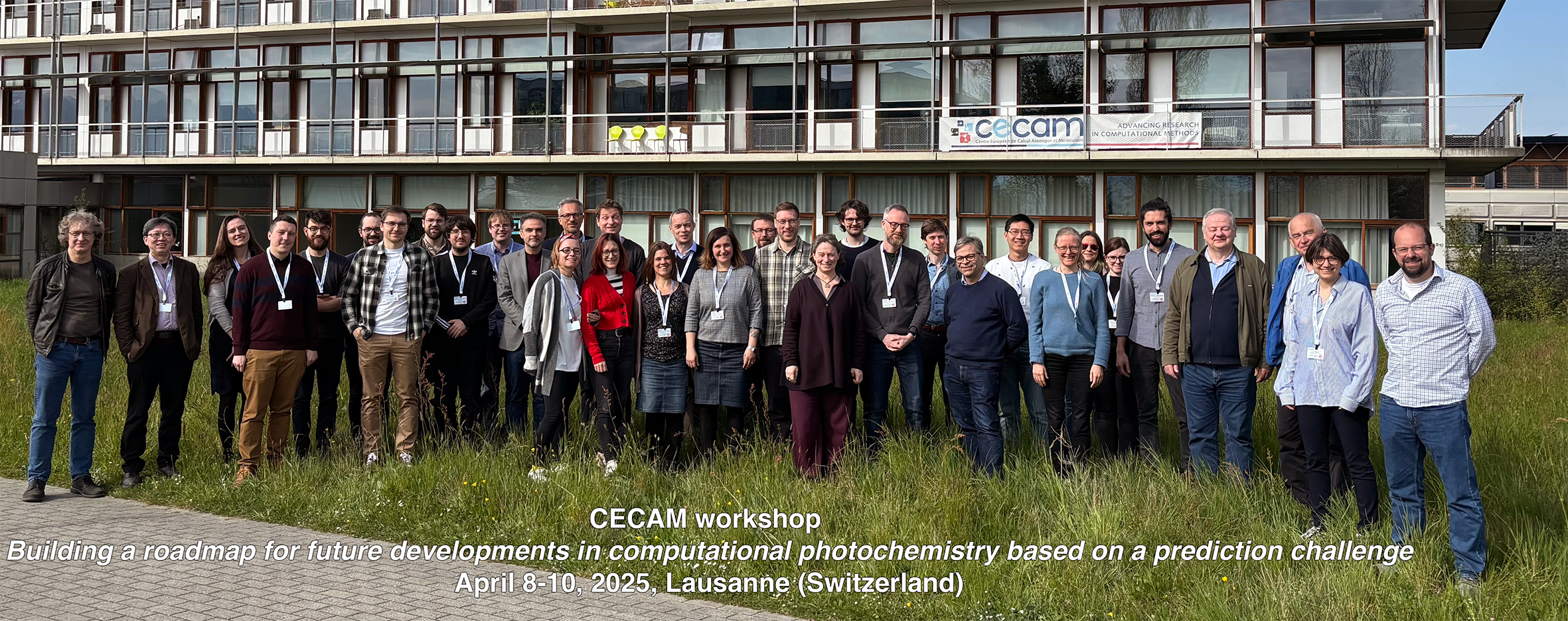}
\caption{Group photograph taken during the CECAM workshop 'Building a roadmap for future developments in computational photochemistry based on a prediction challenge', April 8-10, 2025, Lausanne (Switzerland). Credit: Aude Merola.}
\label{fig:grouppict}
\end{figure}

This Perspective aims to complement the contributions to this Special Topic \textit{Prediction Challenge: Cyclobutanone Photochemistry} by discussing and contrasting their approaches and results for the four main categories highlighted above: photoexcitation, electronic structure, nonadiabatic dynamics, and calculation of observables. The analysis and arguments presented here emanate from a CECAM workshop that took place in Lausanne from the 8th to the 10th of April 2025 and brought together nearly all research groups that participated in this exercise (see Fig.~\ref{fig:grouppict} for a photograph of the participants). The goal of the present work is, therefore, to compare and contrast the various strategies deployed using solely the submitted contributions (and no subsequent work by the contributors or others). We highlight the strengths and weaknesses of various methods that were factually identified during the CECAM workshop, and we suggest, for each topic, a to-do list of calculations and simulations needed to build further conclusions on specific topics. Our analysis naturally leads to more general conclusions about the requirements for the field of computational photochemistry to ensure more accurate, precise, efficient, and reproducible nonadiabatic molecular dynamics simulations in the future, building confidence for other predictions. The reader should keep in mind when reading the various contributions to this Special Topic and the present Perspective that the contributors had a very limited amount of time, 6 months in total (or even less for certain contributors) to (1) study the photochemistry of cyclobutanone from scratch, (2) predict time-resolved experimental observables from their nonadiabatic dynamics simulations, and (3) submit an article to the Journal of Chemical Physics. This Perspective also reflects on the design and realization of this challenge and what could be improved for future iterations, as well as what could be achieved in such a short amount of time.

\section{Analysis of the contributions}
\label{sec:challenge}

In the following, we present the outcome of the discussions on the four main topics (or components) that define each contribution to the challenge: photoexcitation, electronic structure, nonadiabatic dynamics, and calculation of observables. For each section, a dedicated table offers a summary of the main parameters and strategies employed for each contribution. We stress that this Perspective article does not intend to offer a detailed explanation of all the methods that were used by the different participants -- the reader can refer to each contribution where this information (and corresponding references) is available. Yet, we present below a brief overview of the various methods deployed for the nonadiabatic dynamics and the electronic structure, introducing as well their abbreviations.

Electronic-structure methods used for the challenge comprise single-reference methods such as MP2 (second-order Møller–Plesset theory), ADC(2) (algebraic diagrammatic construction to second order) or EOM-CCSD (equation-of-motion coupled cluster singles doubles), and density-based approaches such as DFT (density functional theory) and LR-TDDFT/TDA (linear-response time-dependent density functional theory with or without the Tamm-Dancoff approximation). These single-reference methods only account for dynamic correlation. Methods that effectively only incorporate static correlation and used in this challenge include SA-CASSCF (state-averaged complete active space self-consistent field), MCSCF (multiconfigurational self-consistent field), and FOMO-CASCI (floating occupation molecular orbital complete active space configuration interaction). These methods will collectively be labeled as 'MCSCF' in the following. Some contributions used electronic-structure methods incorporating both static and dynamic correlation, specifically MRCIS (multireference configuration interaction singles) and XMS-CASPT2 (extended multi-state complete active space second order perturbation theory). In the following, these methods including both static and dynamic correlation will be labeled with 'MRPT/MRCI'. The semiempirical approach MRCI/ODM3 (multireference configuration interaction with the semiempirical orthogonalization and dispersion corrections) was also tested. 

A wide variety of approaches were used for the nonadiabatic molecular dynamics. The MCTDH (multiconfiguration time-dependent Hartree) method for quantum dynamics was employed in two contributions with linear and quadratic vibronic coupling (LVC and QVC, respectively) model Hamiltonians. Various methods based on a representation of nuclear wavefunctions with trajectory basis functions were used: DD-vMCG (direct-dynamics variational multiconfigurational Gaussian), AIMC (ab initio multiple cloning), and AIMS (ab initio multiple spawning). Finally, the most commonly applied family of methods for nonadiabatic dynamics was mixed quantum/classical approaches. Trajectory surface hopping (TSH) was deployed in two flavors: FSSH (fewest-switches surface hopping)\cite{tully1991nonadiabatic} or MASH (mapping-approach to surface hopping).\cite{richardson2025mash} Last but not least, an ab initio version of Ehrenfest dynamics with a collapse to a block (TAB)\cite{eschTAB2020} strategy was also represented in this challenge.

\subsection{Photoexcitation}
\label{sec:ic}

\begin{figure}[ht]
\centering
\includegraphics[width=0.6\textwidth]{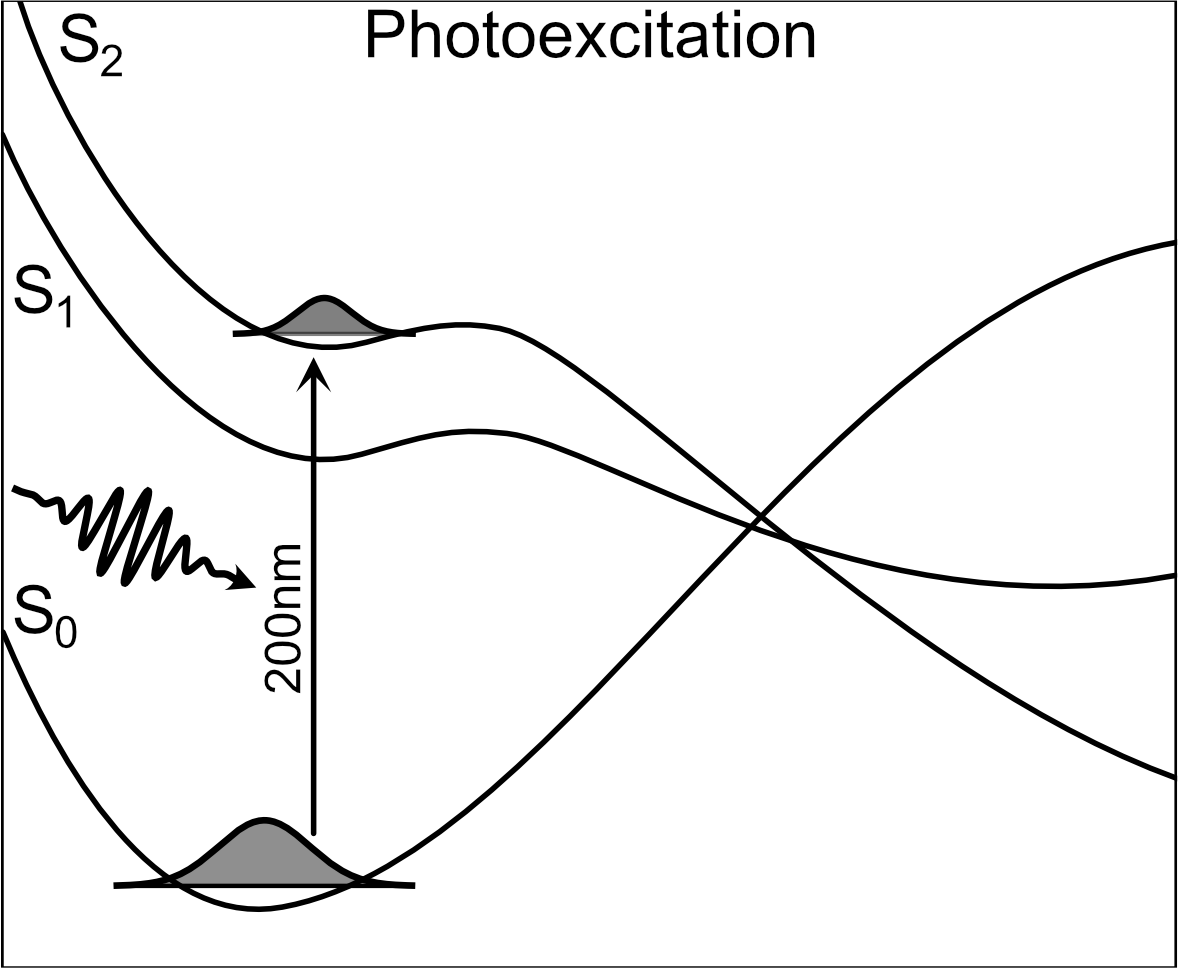}
\caption{Schematic representation of the photoexcitation process.}
\label{fig:schemeic}
\end{figure}

\begin{table}[htp]
\centering
\caption{Selected details about the different strategies for the photoexcitation process.  }
\begin{threeparttable}
\scalebox{0.85}{
\begin{tblr}{
  colspec = {
  |l l| X[1.2,l] X[0.8,l] X[1.6,l] |
  },
  hlines,
  cells = {font=\small} 
}
\SetCell[c=2]{c}  & & \SetCell[c=3]{c} \textbf{Photoexcitation} \\ 
\textit{Label} & \textit{Ref} & \textit{S$_0$} \textit{Sampling} & \textit{Selection} & \textit{Electronic structure}\tnote{\dag}  \\ 
Paper 1$^\otimes$ & \citenum{parker2024cyclo} & Boltzmann (260 K) & No window & DFT/RIJ-TPSS-D3(BJ)/def2-SVPD \\ 
Paper 2 & \citenum{gonzalez-vazquez2024cyclo} & Harmonic Wigner (v=0) & Energy window & MP2/aDZ \\ 
Paper 3$^\otimes$ & \citenum{martinez2024cyclo} & Harmonic Wigner (v=0) & Energy window + weighting & CCSD/aDZ \\ 
Paper 4 & \citenum{lan2024cyclo} & Harmonic Wigner (v=0) & Energy window & DFT/B3LYP/6-31G* \\ 
Paper 5 & \citenum{kirrander2024cyclo} & Harmonic Wigner (v=0) & Energy window & SA3-CASSCF(12/12)/aDZ \\ 
Paper 6 & \citenum{richardson2024cyclo} & Harmonic Wigner (325 K), 1D quantum for puckering mode & No window & DFT/B3LYP/def2-TZVP(0.96 mode rescaling) \\ 
Paper 7$^\otimes$ & \citenum{worth2024cyclo} & Harmonic ground-state vibrational wavefunction & No window & CCSD/6-311++G** and SA3-CASSCF(8/10)/6-311++G** \\ 
Paper 8 & \citenum{garavelli2024cyclo} & Harmonic Wigner (v=0, 7 modes removed) & Energy window and no window & MP2/6-31++G** \\ 
Paper 9 & \citenum{shalashilin2024cyclo} & Harmonic Wigner (v=0) & No window & SA3-CASSCF(12/12)/aDZ \\ 
Paper 10$^\otimes$ & \citenum{barbatti2024cylo} & Positions: Harmonic Wigner (v=0). Momenta: restricted to ZPE & Energy window & SA3-MCSCF(14/14) \\
Paper 11$^\otimes$ & \citenum{penfold2024cyclo} & Harmonic Wigner (100 K) & No window & PBE0/aDZ \\ 
Paper 12 & \citenum{curchod2024cyclo} & Quantum-thermostat BOMD (298 K) & Energy window & MP2/DZ \\ 
Paper 13 & \citenum{levine2024cyclo} & Harmonic Wigner (v=0) & No window & DFT/B3LYP/6-311++G** \\ 
Paper 14 & \citenum{mitric2024cyclo} & Harmonic Wigner (50 K, low-frequency mode removed) & No window & SA(3S,2T)-CASSCF(6/6)/MRCIS/ aDZ(C,O),DZ(H) \\ 
Paper 15 & \citenum{gomez2024cyclo} & Harmonic Wigner (v=0) & Energy window & SA6-CASSCF(8/11)/aDZ \\ 
\end{tblr}
}
\begin{minipage}{0.7\textheight}
\small
\begin{tablenotes}
\item[$\otimes$] Contributions that reported multiple results for the prediction. Only one is reported in this Table. 
\item[\dag] DZ: cc-pVDZ; aDZ: aug-cc-pVDZ; d-aDZ: d-aug-cc-pVDZ.
\end{tablenotes}
\end{minipage}
\end{threeparttable}
\label{tab:IC}
\end{table}

One of the  first steps in preparing a nonadiabatic molecular dynamics simulation is the description of the photoexcitation process (Fig.~\ref{fig:schemeic}). Several approaches for nonadiabatic molecular dynamics could in principle incorporate the pump laser pulse explicitly. Yet, the most commonly applied strategy is to treat the photoexcitation process implicitly -- all contributions in this Special Topic used this approach. In this approach, the photoexcitation process is split into two steps:\cite{janos2025selecting} description of the system on the ground electronic state and its projection onto a given excited electronic state to describe the molecular state resulting from photoexcitation. For trajectory-based methods, the usual protocol requires that pairs of nuclear positions and momenta, coined 'initial conditions', are sampled from a given ground-state distribution and used to initiate trajectories in the selected excited electronic state(s). For methods like MCTDH, multi-layer (ML-) MCTDH, and DD-vMCG, an initial nuclear wavefunction is obtained for the ground electronic state and projected onto the excited state of interest to initiate the dynamics, sometimes after being multiplied by the transition dipole moment connecting the ground electronic state to the excited state of interest.

The majority of contributions making use of trajectory-based approaches for nonadiabatic dynamics (TSH-based methods, AIMS, and AIMC) used a harmonic Wigner distribution for sampling their initial conditions (Table~\ref{tab:IC}). The harmonic Wigner distribution was constructed from the optimized ground-state geometry of cyclobutanone and corresponding harmonic frequencies. Paper 10 used an alternative approach for its sampling: nuclear momenta were selected such that the total energy of each initial condition matches the zero-point energy. Paper 1 sampled the ground-state microcanonical distribution with ab initio molecular dynamics with an average temperature of 260 K. 

Papers 6, 7, 8, 11, 12 discussed the ground-state geometry of cyclobutanone in more detail, highlighting that the molecule exhibits a ring-puckering geometry (its planar structure with C$_{2v}$ symmetry being a transition state in the ground electronic state) with an associated low-frequency mode.  This observation may question the validity of using a harmonic approximation for this mode, given the double-well nature of the ground-state potential energy surface along this mode coordinate, with a barrier estimated from experiment\cite{Scharpen1968} to be $\sim 0.1$ kJ/mol above the two symmetric minima. Paper 12 showed the sensitivity of the ring-puckering dihedral angle and its associated vibrational wavenumber to the level of theory employed. Paper 6 used a hybrid sampling approach combining a sampling of the harmonic Wigner distribution for all normal modes but the ring-puckering one, for which the distribution was described exactly using a 1D quantum calculation, with an overall temperature of 325 K. Paper 8 excluded 7 normal modes from the harmonic Wigner distribution: the low-frequency mode as well as 6 other ones appearing above 3000 cm$^{-1}$, the latter modes being removed to minimize zero-point energy leakage in the subsequent dynamics. Paper 12 used Born--Oppenheimer molecular dynamics (BOMD) combined with a quantum thermostat (QT) to produce a ground-state distribution mimicking the quantum one at 298 K, explictly accounting for the anharmonicity of the normal modes.

Contributions using MCTDH simulations built model potentials either using a QVC Hamiltonian (Paper 11) or a LVC Hamiltonian (Paper 7), with initial conditions either based on a planar (C$_{2v}$) geometry or the ring-puckered one (C$_{s}$) for ground-state cyclobutanone. Paper 7 also used a different set of initial conditions for DD-vMCG calculations, where the initial trajectory basis wavefunctions (TBFs, also called Gaussian wavepackets) were centered around the C$_{2v}$ ground-state geometries. 

The second step in the description of a photoexcitation process is the consideration of the role played by the (pump) laser pulse at 200 nm on the (ground-state) initial conditions. A simple approach consists of applying an energy-windowing restriction to the selection of nuclear coordinates/momenta pairs, mimicking the energy bandwidth of the pump laser pulse employed in the experiment. Some contributions included such a windowing, based on calculations of a photoabsorption cross-section. For example, Paper 2 used the laser bandwidth to select initial conditions; Paper 3 used a weighting of each initial condition by a function proportional to the oscillator strength to account for the effect of the laser pulse; Paper 4 defined a 4-nm window around the central frequency of the laser pulse; Paper 5 a 5-nm window; Paper 10 applied two windows (depending on the level of electronic-structure theory); Paper 12 applied an energy window reproducing the experimental bandwidth of the pump pulse, and Paper 15 used a windowing between 191 and 210 nm.  Paper 8 looked at the effect of the windowing by comparing the excited-state dynamics resulting from 100 initial conditions sampled from a Wigner distribution (no windowing) or initial conditions from the same Wigner distribution but applying windowing to mimic the pulse (6-6.2 eV). The impact of these initial conditions on the nonadiabatic dynamics will be discussed in Sec.~\ref{sec:namd}. All other contributions do not mention a specific windowing of the initial conditions. Most nonadiabatic dynamics methods working with trajectories and using the adiabatic representation considered that the 200-nm pulse would excite cyclobutanone purely in the adiabatic S$_2$ electronic state. This electronic state exhibits (in the Franck--Condon region) a 3s Rydberg character and will be denoted as S$_2(3s)$ in the following (more details in Sec.~\ref{sec:es}).

An interesting practical question related to the selection of initial conditions was raised during the CECAM workshop. What is the impact of sampling the initial conditions using a different level of electronic-structure theory than that used for the nonadiabatic dynamics? The answer to this question is nuanced. While it would make sense to advise that the same level of electronic-structure theory should be used for the initial conditions and the nonadiabatic dynamics, cyclobutanone presents an interesting dilemma. For example, SA-CASSCF does not adequately describe the ring-puckering of cyclobutanone in the ground electronic state. Should calculations then prioritize a better description of the initial conditions at the cost of using a different electronic-structure method, or should they stick to the electronic structure that will be used for the nonadiabatic dynamics? Using a method that describes dynamical correlation for the sampling of initial conditions and switching to SA-CASSCF for the dynamics may lead to artifacts too, as bond lengths are usually longer for SA-CASSCF geometries, inducing a sudden relaxation in the early times of the nonadiabatic dynamics. The group reached the agreement that such a switch should be carefully tested; its impact on the final results is likely to be system dependent, and may more specifically depend on the character of the excited electronic states where the early time dynamics take place. In this sense, the Rydberg state reached upon photoexcitation at 200 nm for cyclobutanone may have been more forgiving. Tests could include a comparison between the distributions of excitation energies obtained from different sets of initial geometries, i.e., sampled with different electronic-structure methods (within a given approach -- harmonic Wigner, QT-BOMD, etc), all using the same electronic structure for the excitation energies. A deviation between distributions may reveal an underlying impact of the sampled ground-state geometries.  A more intensive test would be to perform nonadiabatic dynamics with a given electronic-structure method, but using initial conditions sampled with different electronic-structure methods. Comparison should then be based on the behavior of the trajectories in the early times of the dynamics, observing the 'relaxation' of the molecular structure from one electronic-structure method (used for the ground-state sampling) to another (used for the dynamics) in the bond lengths and possibly on the electronic energies.

These few paragraphs highlight the wide variety of strategies to describe photoexcitation, in particular when defining initial conditions for trajectory-based approaches. The impact of treating the low-frequency mode within a harmonic approximation cannot be directly assessed from the various contributions of the challenge due to subtle variations in protocols. Addressing this impact would, however, constitute a simple follow-up work, as detailed below (Sec.~\ref{sec:ictd}). No contributions included the temporal profile of the pump laser pulse in their simulation in a rigorous way.

\subsubsection{To-do list: Photoexcitation}
\label{sec:ictd}

A simple set of calculations could be performed to test the sensitivity of this cyclobutanone photodissociation prediction to the initial conditions. TSH-based simulations could be conducted with initial conditions sampled from a fully harmonic Wigner distribution, a modified harmonic Wigner distribution to treat the low-frequency mode, and a distribution built from QT-BOMD dynamics -- all initial conditions obtained from the very same level of electronic-structure theory to describe the ground electronic state. The TSH dynamics resulting from these different sets of initial conditions should use the best possible level of electronic-structure theory, but other electronic-structure methods could be tested too to measure the sensitivity of the simulations to the electronic-structure method  used for sampling initial conditions. A large number of initial conditions should be created to allow various schemes of energy-windowing to be tested, measuring more directly their impact on the photochemistry of cyclobutanone and more generally the impact of explicitly including the excitation laser pulse in nonadiabatic dynamics.  

\subsection{Electronic structure}
\label{sec:es}

\begin{figure}[ht]
\centering
\includegraphics[width=0.6\textwidth]{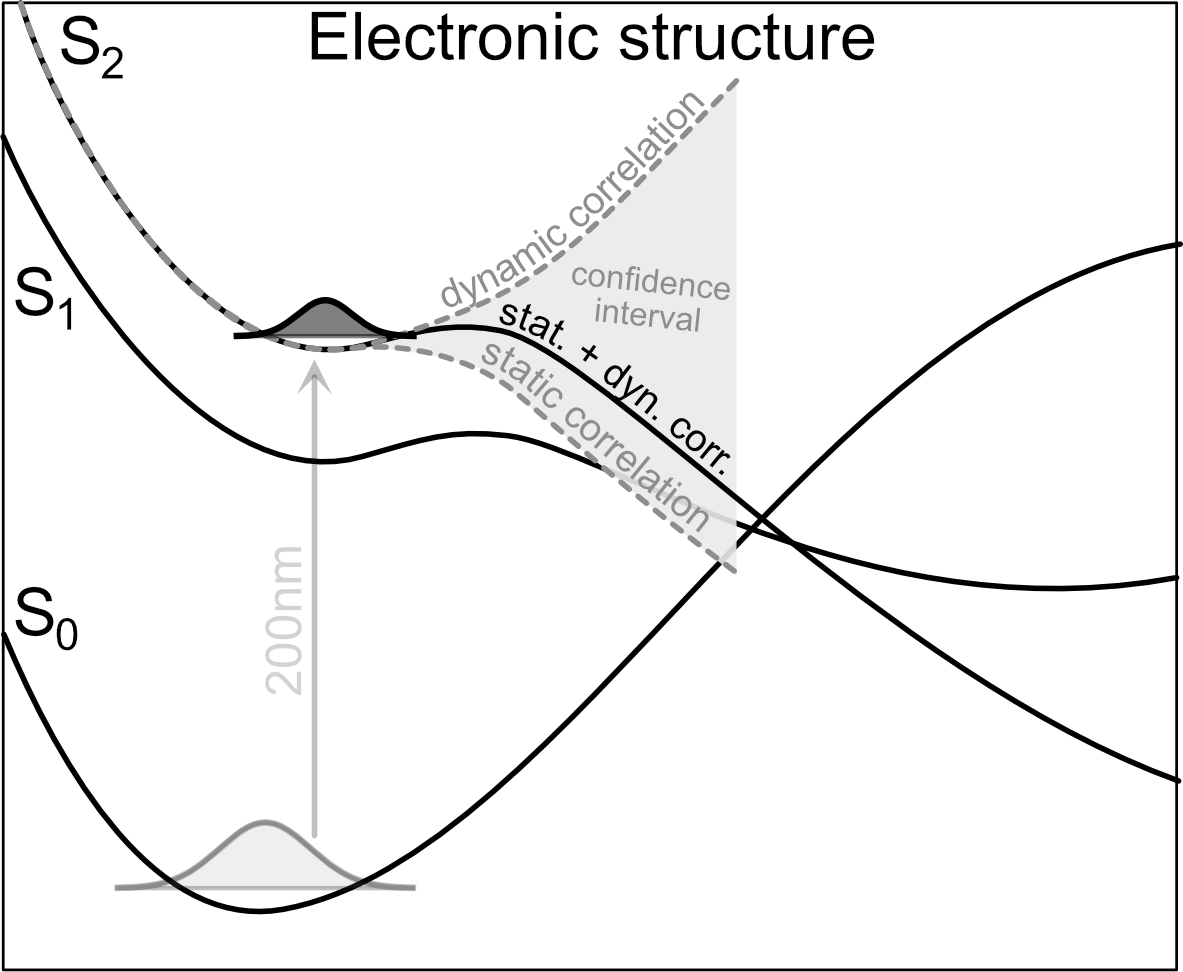}
\caption{Schematic representation of the impact of the level of electronic-structure theory on the potential energy surfaces of cyclobutanone.}
\label{fig:schemeelstr}
\end{figure}

\begin{table}[htp]
\centering
\caption{Selected details about the different strategies for the electronic structure. }
\begin{threeparttable}
\scalebox{0.85}{
\begin{tblr}{
  colspec = {
  |l l| X[1.3,l] X[0.9,l] X[0.7,l] X[0.6,l] |
  },
  hlines,
  cells = {font=\small} 
}
\SetCell[c=2]{c}  & & \SetCell[c=4]{c} \textbf{Electronic structure (ES)} \\ 
\textit{Label} & \textit{Ref} &  \textit{ES method}\tnote{\P} &  \textit{Correlation}& \textit{Basis set}\tnote{\dag} & \textit{ES code} \\
Paper 1$^\otimes$ & \citenum{parker2024cyclo} & LR-TDDFT/TDA/RIJK-PBE0-D3(BJ) & Dynamic & def2-TZVPD & Turbomole \\
Paper 2 & \citenum{gonzalez-vazquez2024cyclo} & SpinA-XMS(4S,4T)-CASPT2(8/8) & Static + dynamic & aDZ & OpenMolcas \\
Paper 3$^\otimes$ & \citenum{martinez2024cyclo} &  EOM-CCSD & Dynamic & aDZ & eT \\
Paper 4 & \citenum{lan2024cyclo} & XMS3-CASPT2(10/8) & Static + dynamic & def2-SVPD & Bagel \\
Paper 5 & \citenum{kirrander2024cyclo} &  SA3-CASSCF(12/12) & Static & aDZ & OpenMolcas \\
Paper 6$^\otimes$ & \citenum{richardson2024cyclo} & SA3-CASSCF(12/11) followed by SS-CASSCF(12/11) AIMD in S$_0$ when encountering instabilities & Static & aDZ & Molpro \\
Paper 7$^\otimes$ & \citenum{worth2024cyclo} & SA3-CASSCF(8/10) & Static & 6-311++G** & Molpro \\
Paper 8 & \citenum{garavelli2024cyclo} & SpinD-XMS(3S,3T)-CASPT2(8/8) &  Static + dynamic & 6-31++G* & OpenMolcas \\
Paper 9 & \citenum{shalashilin2024cyclo} &  SA3-CASSCF(12/12) & Static & aDZ & Molpro \\
Paper 10$^\otimes$ & \citenum{barbatti2024cylo} &  SA3-MCSCF(14/14) & Static & H: DZ; C: aDZ; O: d-aDZ & Columbus \\
Paper 11$^\otimes$ & \citenum{penfold2024cyclo} &  LR-TDDFT/TDA/PBE0 & Dynamic & aDZ & Turbomole \\
Paper 12 & \citenum{curchod2024cyclo} & XMS3-CASPT2(8/8) (followed by MP2/aDZ AIMD in S$_0$) & Static + dynamic  & aDZ & Bagel \\
Paper 13 & \citenum{levine2024cyclo} &  FOMO-CASCI(8/11) & Static & 6-311+G* & TeraChem \\
Paper 14 & \citenum{mitric2024cyclo} & SpinA-SA(3S,2T)-CASSCF(6/6)/MRCIS & Static + dynamic & H: DZ; C/O: aDZ & Columbus \\
Paper 15$^\otimes$ & \citenum{gomez2024cyclo} &  SA6-CASSCF(8/11) & Static & aDZ & OpenMolcas \\
\end{tblr}
}
\begin{minipage}{0.7\textheight}
\small
\begin{tablenotes}
\item[\P] SpinA: spin-adiabatic electronic basis; SpinD: spin-diabatic electronic basis.
\item[$\otimes$] Contributions that reported multiple results for the prediction. 
\item[\dag] DZ: cc-pVDZ; aDZ: aug-cc-pVDZ; d-aDZ: d-aug-cc-pVDZ.
\end{tablenotes}
\end{minipage}

\end{threeparttable}
\label{tab:ES}
\end{table}

A wide variety of electronic-structure methods has been tested during this challenge, as clearly visible from Table~\ref{tab:ES} and schematically depicted in Fig.~\ref{fig:schemeelstr}. The discussion in the following section is articulated around the different families of electronic-structure approaches used for the excited-state dynamics, and comparisons will be based on static calculations and photochemical reaction pathways (the impact of the electronic-structure method on the dynamics will be saved for Sec.~\ref {sec:namd}). We also stress that many participants benchmarked various levels of electronic-structure results, using static calculations. 

Let us begin with contributions based on single-reference or density-functional methods. EOM-CCSD with an aug-cc-pVDZ basis set was used by Paper 3, while LR-TDDFT was used with the following combination of functional and basis set and with/without TDA: TDA/PBE0-D3(BJ)/def2-SVPD and TDA/PBE0-D3(BJ)/def2-TZVPD (Paper 1), TDA/LRC-$\omega$PBE/aug-cc-pVDZ (Paper 3), CAM-B3LYP/cc-pVDZ(H),aug-cc-pVDZ(C),d-aug-cc-pVDZ(O) (Paper 10, not tabulated), TDA/PBE0/aug-cc-pVDZ (Paper 11), or TDA/B3LYP-D3/6-31+G$^{\ast\ast}$ (Paper 15, not tabulated). Papers 10 and 11 tried ADC(2), yet these attempts were unsuccessful -- this method was reported to exhibit artificial, low-energy crossing seams between S$_1$ and S$_0$ for carbonyl-containing molecules.\cite{D1CP02185K} Paper 10 also tried CC2 without success, experiencing convergence issues as detailed earlier in Ref.~\citenum{plasser2014surface}.

Other contributions made use of MCSCF methods for their dynamics. In the following, we mention the active-space size (and state averaging procedure) -- the reader can consult the corresponding references for details about the orbitals included. SA-CASSCF was used in several contributions with various combinations of active spaces and state averaging procedures (with or without triplet states included): SA3-CASSCF(12/12)/aug-cc-pVDZ (Paper 5 and Paper 9), SA3-CASSCF(12/11)/aug-cc-pVDZ followed by SS-CASSCF(12/11)/aug-cc-pVDZ for ground-state dynamics when encountering instabilities (Paper 6), SA3-CASSCF(8/10)/6-311++G$^{\ast\ast}$ (Paper 7, for DD-vMCG with a database, see Sec.~\ref{sec:namd}), and SA6-CASSCF(8/11)/aug-cc-pVDZ (Paper 15). 

MRPT/MRCI methods were represented predominantly by XMS-CASPT2: XMS(4S,4T)-CASPT2(8/8)/aug-cc-pVDZ (Paper 2), XMS3-CASPT2(8/10)/def2-SVPD (Paper 4), CASPT2 corrected SA(3S,3T)-CASSCF(10/8)/6-311++G$^{\ast\ast}$ for linear vibronic coupling model used with ML-MCTDH (Paper 7), XMS(3S,3T)-CASPT2(8/8)/6-31++G$^\ast$ (Paper 8), XMS3-CASPT2(8/8)/aug-cc-pVDZ combined with MP2/aug-cc-pVDZ for ground-state dynamics when dissociation occurred (Paper 12).

Finally, some contributions utilized alternative approaches. Paper 14 used MRCIS based on SA(3S/2T)-CASSCF(6/6)/cc-pVDZ(H),aug-cc-pVDZ(C,O), Paper 10 tried a split MCSCF approach containing five perfect pairings (PPs, each with two electrons in two orbitals) and one complete active space with four electrons and four orbitals, resulting in a SA3-MCSCF(14/14)/cc-pVDZ(H),aug-cc-pVDZ(C),d-aug-cc-pVDZ(O), and Paper 13 employed TD-CASCI at the FOMO-CASCI(8/11)/6-311+G$^\ast$ level of theory. Paper 10 also tested an MRCI approach based on the semiempirical ODM3 method.

Let us discuss the performance of this zoo of electronic-structure methods for the different regions of nuclear configuration space visited by cyclobutanone upon photoexcitation to the S$_2(3s)$ electronic state. Most methods, including single-reference and LR-TDDFT ones, appear to offer an accurate description of S$_2(3s)$ in the Franck--Condon region. The minimum-energy structure for this electronic state is close to the Franck--Condon region, with small geometrical variations from the ground-state structure, namely its planarization. Different groups benchmarked (in the Franck--Condon region) the quality of their chosen strategy with respect to basis set and also higher-level electronic-structure methods like CC3 (Papers 3 and 12) or adaptive and Monte Carlo configuration interaction (Paper 5). The adequate description of the S$_2(3s)$ state was further confirmed by the calculation of photoabsorption cross-sections in good agreement with the experimental one, either using the nuclear ensemble approach (reproducing the overall shape of the S$_2(3s)\leftarrow$S$_0$ band), autocorrelation methods with MCTDH (Paper 7 and Paper 11), or Franck--Condon Herzberg-Teller factors (Paper 12 and Paper 14) -- the two last strategies being able to capture the vibronic progression associated with this transition. 

Isolating critical geometries (minima, transition states, minimum-energy conical intersections) on potential energy surfaces revealed that cyclobutanone can evolve \textit{adiabatically} in S$_2$, i.e., switching its electronic character from Rydberg to valence, with an associated bond elongation (carbon-carbon bond next to the carbonyl group) leading eventually to ring opening and nonadiabatic transitions from S$_2$ to S$_1$ first, and then to S$_0$. The adiabatic evolution away from the region of Rydberg character involves a transition state, whose barrier height with respect to the S$_2(3s)$ minimum (Fig.~\ref{fig:schemeelstr}) is critical for the timescales depicted by nonadiabatic dynamics simulations (see Sec.~\ref{sec:namd}). Perhaps more importantly, moving away from the Franck--Condon region revealed the limitations of single-reference methods and LR-TDDFT. Different contributions used interpolated or intrinsic-coordinate pathways between critical geometries to compare electronic energies along these paths as obtained by single-reference (and LR-TDDFT) and MCSCF or MRPT/MRCI methods. Such a benchmark beyond the Franck--Condon region highlights the importance of static correlation and, as such, the potential issues that EOM-CCSD or LR-TDDFT can have beyond the description of the initially formed Rydberg state (Paper 12). But these pathways also make explicit the dramatic impact played by dynamic correlation on the S$_2$ barrier controlling the adiabatic evolution from Rydberg to valence character (Paper 5, Paper 6, Paper 8, Paper 12). SA-CASSCF calculations predict an adiabatic evolution in S$_2$ that is essentially barrierless, whereas including dynamic correlation (for example with XMS-CASPT2) leads to the appearance of a small yet non-negligible barrier (certain contributions isolated a transition state).  

Another important parameter affecting the height of this barrier in S$_2$ is the choice of the basis set -- the barrier height is larger by a few kJ/mol with XMS(3S,3T)-CASPT2(8/8)/6-31++G$^\ast$ (14.5 kJ/mol, Paper 8) than with XMS3-CASPT2(8/8)/aug-cc-pVDZ (18.3 kJ/mol, Paper 12). Paper 3 highlighted the important variations in energy observed with basis sets like def2-SVPD or 6-31+G$^\ast$, in comparison to aug-cc-pVDZ. These findings are corroborated by the results of Paper 6, showing significant differences in photoproduct yields between aug-cc-pVDZ and 6-31+G$^\ast$, using SA3-CASSCF(12/11). Yet, it should be pointed out that the active-space orbitals for the (12/11) active space with 6-31+G$^\ast$ appear imbalanced, i.e., exhibiting asymmetries, in comparison to the active-space orbitals using aug-cc-pVDZ (Supplementary Material of Paper 6). The impact of this imbalance in the active-space orbitals (also observed in the active space of Paper 4) cannot be disentangled from the impact of the basis set on the potential energy surfaces without further calculations. 

For the original pathway involving ring opening via the adiabatic path in S$_2$, cyclobutanone (in a diradical open form) reaches a conical intersection with S$_1$. XMS-CASPT2 appears to bring this intersection close in energy to S$_0$, while SA-CASSCF shows a larger energy splitting between S$_0$ and the other excited electronic states (Paper 12). 

Besides this adiabatic pathway, another path was located in Paper 12, involving the passage through a S$_2$/S$_1$ intersection characterized by a closed cyclobutanone (the structure of the corresponding minimum-energy conical intersection was also identified in Paper 5) and leading to the ring opening of the molecule in S$_1$. 

Overall, the photochemistry of cyclobutanone initiated by a laser pulse at 200 nm offers a fantastic playground for testing electronic-structure strategies for various processes. Upon photoexcitation, the molecule undergoes a photophysical process by relaxing in S$_2$ with Rydberg (3s) character, leading to the appearance of vibronic progressions in the photoabsorption cross-section of cyclobutanone. This process is accurately captured by single-reference and MRPT/MRCI methods that incorporate dynamic correlation, as long as an adequate basis set is employed. However, cyclobutanone can escape this S$_2(3s)$ minimum via a transition state, leading to an adiabatic change of electronic character and triggering its rich photochemistry. This second process is far more demanding for electronic-structure methods. Not only should the method be able to describe bond-breaking processes (via the inclusion of static correlation) but it should capture the barrier height properly to adequately describe the timescale for the adiabatic process to take place. The contributions to this challenge allow us to highlight the critical role of the electronic-structure method (in particular, the inclusion of both static and dynamic correlation to describe the adiabatic evolution in S$_2$) and the basis set, but more calculations are required to fully characterize the barrier on this adiabatic path. While comparison of experimental and calculated lifetimes for specific features in the MeV-UED observables (see Sec.~\ref{sec:obs} below) may offer an indirect indication that specific approaches (in particular, XMS-CASPT2 with a large basis set) describe this barrier adequately, the uncertainty related to this barrier height in S$_2$ can only be reduced by additional calculations at a higher level of theory, possibly including CC3 if static correlation at this nuclear configuration is not too important. Discussions during the CECAM workshop pointed out that larger active spaces, in particular when combined with a perturbation-based description of dynamic correlation, may not provide a better estimate of the barrier height (see for example the SI of Ref.~\citenum{levine2009} for a discussion on active-space size with SA-CASSCF).

This analysis and the discussions occurring during the CECAM workshop allow us to provide the following practical guidance for future similar works: (i) It is absolutely critical to benchmark electronic-structure methods beyond the Franck--Condon region (by optimizing critical points on the potential energy surfaces and interpolated pathways) to detect potential issues before conducting a nonadiabatic molecular dynamics simulation. For cyclobutanone, these issues are the limitation of single-reference methods or LR-TDDFT to describe the ring opening of cyclobutanone, the shortcomings of SA-CASSCF in describing the barrier along the adiabatic pathway in S$_2$, or the sensitivity of the overall process to the basis set employed. Furthermore, such benchmarks are critical to test the stability of different active spaces beyond the Franck--Condon region -- a key requirement for nonadiabatic dynamics; (ii) Any work reporting an active-space method should report, at a minimum, a graphical depiction of the active orbitals used, but ideally also readable files (e.g., a molden file) containing the active-space orbitals for various critical geometries explored; (iii) Small differences in the electronic-structure method (state averaging, specific orbitals present in an active space of a given size, apparently similar basis sets) can lead to a significant difference in the description of electronic energies in and beyond the Franck--Condon region -- it is crucial for the community to fully report all the parameters used for a given calculation.  

\subsubsection{To-do list: Electronic structure}
\label{sec:estd}

The barrier along the adiabatic pathway in S$_2$ should be investigated with various electronic-structure methods, possibly starting from the molecular geometries obtained with XMS-CASPT2 and comparing electronic energies along the path connecting the S$_2$ minimum to the transition state. Calculations with CC3 and a large basis set would provide a first estimate of the barrier height. Having this reference (or energies from another high-level electronic-structure method) would support the determination of an adequate active space and basis set to describe the energy barrier with XMS-CASPT2, which can then be used to reoptimize the geometries of the minimum and the transition state in S$_2$. Another element that would deserve attention is the accuracy of the different electronic-structure methods to describe the intersection seams between electronic states (and not just the minimum-energy conical intersections).

As detailed in the previous section, most of the methods used for the electronic structure are 'well-established' and it would be highly beneficial for the community to test and uncover the potential of newer methods -- e.g., MR-SF-TDDFT,\cite{lee2018mrsftddft} (variational\cite{scott2024vmcpdft}) MC-PDFT,\cite{manni2014mcpdft} rotated multistate CASPT2 variant,\cite{battaglia2020xdwcaspt2,battaglia2021rmscaspt2} or multireference-ADC\cite{sokolov2018mradc} -- and possibly stimulate the implementation of necessary electronic properties (nonadiabatic coupling vectors, nuclear gradients, observables). 

Connected to this last point, the nonadiabatic dynamics community should reinforce its interactions with developers of electronic-structure methods and use such pathological cases as tests of state-of-the-art high-level methods beyond the Franck--Condon region. This challenge also highlights the need for researchers spanning the boundary of nonadiabatic molecular dynamics and electronic-structure theory.

\subsection{Nonadiabatic dynamics}
\label{sec:namd}

\begin{figure}[ht]
\centering
\includegraphics[width=0.6\textwidth]{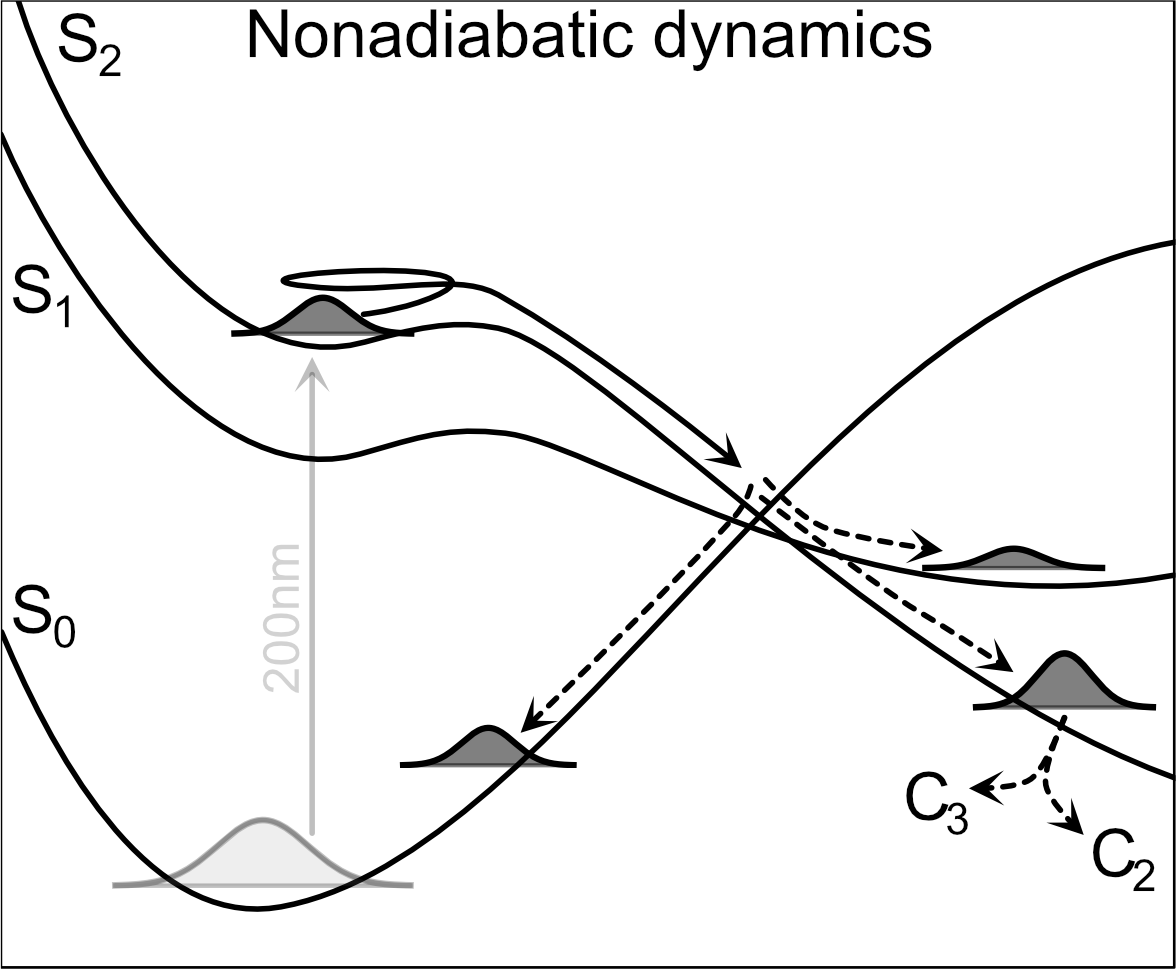}
\caption{Schematic representation of the nonadiabatic dynamics processes related to cyclobutanone.}
\label{fig:schemenamd}
\end{figure}

\begin{table}[htp]
\centering
\caption{Selected details about the different strategies for the nonadiabatic dynamics. MQC=mixed quantum/classical, TBF=trajectory basis functions.}
\begin{threeparttable}
\scalebox{0.90}{
\begin{tblr}{
  colspec = {
  |l l| X[1.0,l] X[1.0,l] X[0.7,l] X[1.0,l] X[0.6,l] X[0.7,l] |
  },
  hlines,
  cells = {font=\small} 
}
\SetCell[c=2]{c}  & & \SetCell[c=6]{c} \textbf{Nonadiabatic dynamics} \\ 
\textit{Label} & \textit{Ref} & \textit{\# trajectories} & \textit{Dynamics method} & \textit{Family} & \textit{Code} & \textit{Time step (fs)} & \textit{Total time (ps)} \\
Paper 1$^\otimes$ & \citenum{parker2024cyclo} & 100 & FSSH & MQC & Turbomole & 0.48 & 3 \\
Paper 2 & \citenum{gonzalez-vazquez2024cyclo} & 300 & DC-FSSH & MQC & SHARC & 0.5 & 0.7 \\
Paper 3$^\otimes$ & \citenum{martinez2024cyclo} & 18 & AIMS & TBF & FMS90 & 0.5 & 0.5 \\
Paper 4 & \citenum{lan2024cyclo} & 100 & DC-FSSH & MQC & JADE-NAMD & 0.5 & 0.5 \\
Paper 5 & \citenum{kirrander2024cyclo} &  229 & MISH & MQC & SHARC-devel & 0.5 & 0.3 \\
Paper 6 & \citenum{richardson2024cyclo} &  198 & unSMASH & MQC & in-house code & 0.5 & 0.5 \\
Paper 7$^\otimes$ & \citenum{worth2024cyclo} &  512 TBFs $\quad$ 32 (12D) $\otimes$ 16 (14D) & DD-vMCG & TBF & Quantics & variable (ca 0.01) & 0.5 \\
Paper 8 & \citenum{garavelli2024cyclo} & 100 & DC-FSSH & MQC & COBRAMM & 1 & 2 \\
Paper 9 & \citenum{shalashilin2024cyclo} & 39 & AIMC & TBF & AIMC-devel & 0.06 & 0.2 \\
Paper 10$^\otimes$ & \citenum{barbatti2024cylo} & 93 & DC-FSSH & MQC & Newton-X & 0.5 & 10 \\
Paper 11$^\otimes$ & \citenum{penfold2024cyclo} & 298 & DC-FSSH & MQC & Newton-X & 0.5 & 5 \\
Paper 12 & \citenum{curchod2024cyclo} & 119 & DC-FSSH & MQC & ABIN & 0.25 & 2 \\
Paper 13 & \citenum{levine2024cyclo} & 150 & TAB & MQC & TAB-DMS & 0.12 & 0.25 \\
Paper 14 & \citenum{mitric2024cyclo} & 52 & FSSH & MQC & FISH & 0.2 & 0.36 \\
Paper 15 & \citenum{gomez2024cyclo} & 294 & DC-FSSH &  MQC & SHARC & 0.5 & 0.25 \\
\end{tblr}
}
\begin{minipage}{0.7\textheight}
\small
\begin{tablenotes}
\item[$\otimes$] Contributions that reported multiple results for the prediction. 
\end{tablenotes}
\end{minipage}
\end{threeparttable}
\label{tab:namd}
\end{table}

Table~\ref{tab:namd} shows the distribution of techniques used to treat the nonadiabatic dynamics (see Fig.~\ref{fig:schemenamd} for a schematic depiction). In short, 12 out of the 15 contributions used (or tested) a mixed quantum/classical approach, 3 contributions employed a technique based on TBFs, and 2 contributions worked with MCTDH.   

FSSH was the most commonly employed flavor among the mixed quantum/classical methods (Papers 2, 4, 8, 10, 11, 12, 14, 15), using an energy-based decoherence correction in all cases except in Papers 1 and 14 (which used none). The total number of trajectories sampled for each contribution is given in Table~\ref{tab:namd}, and we focus here on additional computational details showing the diversity of simulations even for the commonly-employed FSSH. Most simulations were performed with a 0.5 fs time step, except Paper 8 (1 fs), Paper 12 (0.25 fs), and Paper 14 (0.2 fs). Nonadiabatic coupling elements were determined based on an explicit calculation of the nonadiabatic coupling vectors (NACVs, Papers 1, 4, 6, 10, 12, 14), or using time-dependent Baeck-An (TDBA, Papers 10, 11) or approaches using overlaps of electronic wavefunctions (Hammes-Schiffer-Tully scheme: Papers 5, 6, 15). Rescaling of nuclear velocities (momenta) following a hopping event was performed along the NACVs (Papers 1, 4, 5, 6, 10, 12, 14), isotropically (Papers 2, 15), or in the direction of the nuclear momenta with a reduced kinetic energy reservoir (Paper 10). 
Two contributions used different MASH based methods:\cite{richardson2025mash} Paper 5 employed the mapping-inspired method of Runeson and Manolopoulos (MISH),\cite{runeson2023mish} whereas Paper 6 used the size-consistent uncoupled-spheres multi-state generalization of the original MASH algorithm (unSMASH). Both MASH-based nonadiabatic dynamics used a time step of 0.5 fs. Paper 12 tested the impact of the nonadiabatic algorithm by running Landau-Zener surface hopping (LZSH) trajectories on a subset of trajectories (for comparison with FSSH). Paper 6 also tested the influence of spin-orbit couplings using the classical path approximation along pre-computed MASH trajectories.
 Paper 13 used a novel ab initio version of Ehrenfest dynamics with the TAB decoherence correction\cite{eschTAB2020} with a time step of 0.12 fs. A few contributions based on a mixed quantum/classical method included spin-orbit coupling to monitor the importance of intersystem crossing processes: Paper 1 used LZSH to estimate the importance of intersystem crossing along the FSSH trajectories, Paper 2 employed a spin-adiabatic representation in FSSH (4 singlets and 4 triplets) and Paper 8 a spin-diabatic representation (3 singlets and 3 triplets), Paper 12 tested the influence of spin-orbit coupling with LZSH (3 singlets and 3 triplets), Paper 14 used a spin-adiabatic approach (3 singlets and 2 triplets), Paper 15 performed a test with a spin-adiabatic representation (12 singlets and 12 triplets). All of the spin-adiabatic approaches neglected contributions to the total nuclear forces originating from the nuclear derivative of the spin-orbit coupling Hamiltonian.\cite{mai2015IJQC} 

The three main approaches using an on-the-fly propagation of TBFs were represented in this challenge. Paper 3 employed AIMS within the saddle-point approximation of order zero to describe the couplings between TBFs and the independent first generation approximation, starting the dynamics with 64 (LR-TDDFT/TDA) or 18 (EOM-CCSD) initial TBFs. Paper 7 employed DD-vMCG within the local harmonic approximation for calculating the matrix elements coupling the TBFs. DD-vMCG made use of a database to collect electronic-structure information and reduce the computational cost of the overall dynamics. Different DD-vMCG runs were performed with various numbers of TBFs to grow the database and points were added to the database, i.e., calculated on-the-fly, whenever a molecular structure had an atom over 0.2 bohr away from any structure present in the database. The final database comprised 22391 structures and the final run was performed with a 32/16 TBF basis set (32 TBFs for the key 12 modes and 16 for the others, resulting in 512 correlated TBFs) without the need to include any additional points in the database but using Shepard interpolation between points when needed. Paper 9 used AIMC, starting the dynamics with 39 Ehrenfest TBFs giving rise to 121 branches due to cloning events (the nuclear time step used was 0.06 fs).

Last but not least, quantum dynamics simulations using (ML-)MCTDH were performed, based on model potentials constructed from a vibronic coupling Hamiltonian parametrized with the level of electronic-structure theory discussed in Section~\ref{sec:es}, both including spin-orbit coupling. Paper 7 used a linear vibronic coupling Hamiltonian with 12 important normal modes, while Paper 11 employed a quadratic coupling Hamiltonian with 8 selected normal modes. The MCTDH calculations in Paper 7 were used to benchmark the electronic-structure theory by comparing theoretical and experimental photoabsorption spectra, rather than to study the dynamics.

The results of the various nonadiabatic molecular dynamics simulations were summarized in most contributions by using time traces of electronic-state (in the adiabatic and diabatic representations) and photoproduct populations, as well as the photoproduct yields at the end of the dynamics. The first (and only) agreement between all contributions is that intersystem crossing does not play a large role in the photodynamics of cyclobutanone following excitation at 200 nm. Other results are somewhat more nuanced. Upon photoexcitation, cyclobutanone reaches a Rydberg state with $3s$ character, and the time spent by the molecule near the minimum of the S$_2(3s)$ state upon immediate relaxation varies greatly with the level of electronic-structure theory, as anticipated in the previous section (and depicted in Fig.~\ref{fig:schemeelstr}). 

Nonadiabatic dynamics simulations based on LR-TDDFT or EOM-CCSD predict that cyclobutanone remains for an extended period of time (up to multiple picoseconds) in the region of configuration space near the S$_2(3s)$ minimum -- an observation in line with earlier MCTDH simulations using 5 modes and based on EOM-CCSD.\cite{Kuhlman2012} In stark contrast, all mixed quantum/classical dynamics simulations conducted with SA-CASSCF predict a very fast departure from the Franck--Condon region (few tens to a hundred femtoseconds) via the adiabatic process in S$_2$ transforming the electronic character of the (adiabatic) electronic state from Rydberg 3s to valence, leading to the ring opening of cyclobutanone and its decay to S$_1$. Mixed quantum/classical MCSCF dynamics (Paper 10) also predicts ring opening, but on a much longer timescale ($\sim$9 ps). In hindsight, the active space captures dissociation but is too small for the valence–Rydberg coupling necessary to describe the adiabatic change of electronic character in S$_2$. Mixed quantum/classical approaches using XMS-CASPT2 predict that the nuclear wavepacket formed in S$_2$ remains near the S$_2(3s)$ minimum for a few hundred femtoseconds, trapped there by the barrier discussed in Sec.~\ref{sec:es}, before evolving adiabatically and undergoing a ring opening. Another pathway away from the Franck--Condon region was also observed, with an intersection seam between S$_2$ and S$_1$ not involving a ring opening (Paper 12) -- the ratio of the adiabatic pathway to this closed pathway was 3:1. 

The steady-state photoabsorption cross-section of cyclobutanone exhibits a broadened vibronic structure for the S$_2(3s)\leftarrow$S$_0$ band, implying that the initial nuclear wavepacket should remain in the Franck--Condon region for a brief period of time ($<$ 1 ps). The vibronic structure of the photoabsorption cross-section is indeed closely reproduced by quantum-dynamics contributions using (ML-)MCTDH, even if the higher resolution of the vibronic progressions attests to a nuclear wavepacket spending an extended amount of time in the Franck--Condon region in these simulations, possibly due to the level of electronic-structure theory used to describe the ring-opening process. MCTDH-based dynamics focused on the early times, until the nuclear wavepacket performed its adiabatic transfer in S$_2$ towards a region of valence electronic character (note the use of diabatic populations in these contributions, hence any comparison of quantities labeled with S$_2$ or S$_1$ notations should be performed with great care between contributions). Paper 11 also noted the challenge of describing large molecular distortions within the vibronic coupling model and of converging the simulation of a hot ground-state nuclear wavepacket. DD-vMCG trajectories (Paper 7), based on a database built from SA-CASSCF electronic-structure calculations (see Sec.~\ref{sec:es}) showed oscillations of cyclobutanone over the first 500 fs of dynamics but with no occurrence of ring opening.

Nonadiabatic molecular dynamics approaches combined with an MCSCF or MRPT/MRCI method (and not restricted by the definition of model Hamiltonian) all showed that, upon the adiabatic transformation of the S$_2$ electronic state from Rydberg to valence character, cyclobutanone ring-opens and reaches an intersection with S$_1$. Focusing on the population traces, it appears that SA-CASSCF-based nonadiabatic dynamics reach a higher S$_1$ population than XMS-CASPT2-based simulations. Interpolated pathways comparing these regions indicate that the intersection seam between S$_2$ and S$_1$ is closer to S$_0$ with XMS-CASPT2 than with SA-CASSCF, for the same active space and basis set (see for example Paper 12). A concern raised during the CECAM workshop was that higher-lying 3p Rydberg states may decrease in energy during the dynamics and along the adiabatic transformation in S$_2$, potentially becoming relevant to the dynamics. In the Franck–Condon region, a large energy gap exists between the absorption band assigned to the S$_2$(3s)$\leftarrow$S$_0$ transition and those characteristic of higher singlet states. Consequently, most studies restricted their simulations to the three lowest singlet states (S$_0$, S$_1$, S$_2$), with the exception of Papers 2, 3, and 15 (Paper 12 analyzed the potential importance of higher excited states in its Supplementary Material). Single-reference approaches such as LR-TDDFT facilitate the inclusion of a larger number of excited states. LR-TDDFT-based nonadiabatic dynamics in Paper 15 revealed a non-negligible S$_3$ population, while Paper 3 also incorporated S$_3$ and observed population transfer to this state, albeit to a much smaller extent. Including additional Rydberg states is particularly demanding for active-space-based methods and, as noted earlier, might be highly sensitive to the choice of basis set. 

Overall, most nonadiabatic simulations observed the formation of similar photoproducts, independently of the electronic-structure method as long as it allowed an adequate description of bond breaking. The photoproducts are mostly formed when the ring-open cyclobutanone reaches the ground electronic state, with the formation of CO + C$_3$ (cyclopropane or propene) or ethene + C$_2$ (ketene). Other processes were also observed, such as the reformation of hot ground-state cyclobutanone. 

Paper 9, using AIMC and SA-CASSCF, also observed a ring-opening process occurring at the $\beta$ \ce{C-C} bond (with respect to the carbonyl group). The authors mentioned that this observation may be coming from the limited sampling due to the time constraint of the challenge. Paper 5, using the very same level of electronic-structure theory but with the MISH approach, did not report such a mechanism. 
Paper 1, using LR-TDDFT/TDA/def2-SVPD and FSSH, also reported a minimum-energy conical intersection structure exhibiting a stretched $\beta$ \ce{C-C} bond, Paper 3 (AIMS/LR-TDDFT/TDA) mentioned the stretching of this bond when the S$_2$ population decays, and Paper 15 (FSSH/LR-TDDFT/TDA) showed a product formed from a $\beta$ \ce{C-C} bond photolysis. 

Some mixed-quantum/classical and AIMS dynamics employing LR-TDDFT(/TDA) for the electronic structure exhibited longer timescales for the population decays and a smaller proportion of photodissociated molecules. In Paper 1 (FSSH and LR-TDDFT/TDA/PBE0-D3(BJ)), only 8\% of trajectories showed a photodissociation after 10 ps of dynamics (results with def2-TZVPD were similar, but with a short timescale of 3 ps). In Paper 3 (AIMS and LR-TDDFT/TDA/LRC-$\omega$PBE), the S$_2$ lifetime was estimated to be at least 2 ps. Conversely, Paper 10 reported LR-TDDFT/CAM-B3LYP FSSH dynamics (using TDBA) with an immediate decay of the S$_2$ population (0.4 ps of lifetime), but a large number of trajectories did not show a ring opening. Similarly, FSSH dynamics using LR-TDDFT/TDA/B3LYP in Paper 15 led to fast decay of the initial S$_2$ population (half life of $\sim$ 100 fs). It is not possible to pinpoint the cause of these differing results at this point, because too many factors vary at once -- initial conditions, the precise level of electronic structure, and the details of the dynamics algorithm.

Concerning the formation of photoproducts in the ground-electronic state, Paper 1 conducted an interesting set of simulations to test the formation of products under 'thermal conditions' by running ground-state (Born--Oppenheimer) ab initio molecular dynamics at high temperatures. Heating the parent cyclobutanone to 4363 K (representative of 5.1 eV according to the authors) leads to the formation of none of the potential photoproducts. 10000 K is required (14.3 eV) to fragment cyclobutanone in the ground electronic state. Last but not least, Paper 8 conducted a test (in its Supplementary Material) of the impact of the laser pulse on the population traces by comparing the results of nonadiabatic dynamics simulations initialized from a full Wigner distribution or a restricted one with an energy-windowing (see Sec.~\ref{sec:ic} for details). The lifetime of the S$_2$ state is larger with the approximate account of the laser pulse bandwidth (1089$\pm$74 fs with the windowing vs 822$\pm$45 fs without), while the photoproduct ratios did not change significantly. 

The total propagation time of the nonadiabatic dynamics simulations varies significantly between contributions: $t\le$ 500 fs (Papers 3, 4, 5, 6, 7, 9, 13, 14, 15),  500 fs $<t\le$ 1 ps (Papers 2, 3, 10), 1 ps $<t\le$ 2 ps (Papers 8, 11, 12), and $t>$ 2 ps (Papers 1, 10). Certain contributions extended the nonadiabatic dynamics in the ground electronic states by using Born--Oppenheimer molecular dynamics (using the last step of the nonadiabatic molecular dynamics as starting point, with initial nuclear momenta and positions given by the nonadiabatic dynamics) combined with a switch of the electronic-structure method. Paper 6 used SS-CASSCF (instead of SA-CASSCF) when convergence issues were encountered in the ground-state dynamics. Paper 12 used Born--Oppenheimer molecular dynamics combined with MP2/aug-cc-pVDZ to propagate further the trajectories that reached the ground electronic state and had already suffered a dissociation. Such variations in the overall simulation times (and the time-dependence of the photoproducts yields) should be kept in mind when comparing the final C$_3$:C$_2$ ratios reported in each article. These variations also reflect the fact that, despite offering an overall proper description of the expected photoproducts, the nonadiabatic dynamics simulations of this challenge differ significantly in the timescales of photoproduct formation and their ratios. This key observation of the challenge will become important in the next Section when discussing the resulting experimental signals predicted.  

The following important practical considerations emerged when discussing the results of the nonadiabatic dynamics simulations during the CECAM workshop.
\begin{enumerate}
\item[(\textit{i})] Even when the same level of electronic-structure theory is \textit{in principle} used, subtle variations in the implementation of the method or its interface with the nonadiabatic dynamics code may result in different results. For example, the outcome of the electronic-structure calculation might be used differently in the nonadiabatic code, whether nonadiabatic couplings are determined from nonadiabatic coupling vectors, wavefunction overlap, or TDBA couplings (or also the use of database with interpolation in DD-vMCG). This observation means that great care should be taken when comparing the results of different nonadiabatic dynamics methods with seemingly the same level of electronic-structure theory, and is reminiscent of earlier works testing the sensitivity of nonadiabatic dynamics to small perturbations.\cite{jirasensitivityNAMD2024} 
\item[(\textit{ii})] Crashed trajectories should be investigated with great care, as they often reveal an underlying electronic-structure problem that needs to be fixed. Simply discarding them may result in a bias of the dynamics. The number of discarded trajectories should always be reported and their impact on the dynamics discussed.
\item[(\textit{iii})] Cyclobutanone offered a very challenging test for nonadiabatic molecular dynamics thanks to the presence of a barrier near the Franck--Condon region -- this activated process in an excited state challenged the electronic-structure methods and highlighted that simulations can obtain meaningful photoproducts (and, in some cases, their ratios) yet with artificially short or long formation lifetimes. A rule of thumb calculation mentioned in the workshop in connection to Paper 8 revealed that an increase of $\sim$ 4 kJ/mol for the barrier means a rate decrease by a factor $\sim 3$. In ground-state chemistry, $\sim$ 4 kJ/mol is considered chemical accuracy, but this value may be different from what is acceptable in terms of 'photochemical accuracy', at least as long as excited-state lifetime and decay kinetic predictions are concerned, effectively leading to uncertainties roughly spanning one order of magnitude. Paper 13 offered similar considerations by applying RRKM theory to understand the impact of the barrier they observed in S$_1$ on the deactivation timescale. The presence of this barrier also underlines the importance of an adequate description of the photoexcitation process and the initial internal energy of the photoexcited molecule (see Sec.~\ref{sec:ic}).
\item[(\textit{iv})] The present results do not allow us to discuss the importance of decoherence corrections in FSSH dynamics. If anything, the results presented in the context of this challenge could indicate that the importance of using decoherence corrections may be correlated with the electronic-structure method used, as the latter may dramatically affect the nonadiabatic dynamics and make issues with overcoherence more or less important. 
\item[(\textit{v})] As with the level of electronic-structure theory (Sec.~\ref{sec:es}), reproducibility of the results requires the users to provide as many details as possible about their simulations: decoherence correction (and its parameters), number of frustrated hops, momentum inversion or not after frustrated hops, strategy for rescaling of velocities, input parameters for the specific implementation used (and its details if not published in a broadly available code), and exact version of the codes. 
\end{enumerate}

\subsubsection{To-do list: Nonadiabatic dynamics}
\label{sec:namdtd}

A more detailed analysis of the differences observed between DD-vMCG and other trajectory- or TBF-based approaches could be obtained by running all the methods with the DD-vMCG database employed for its final production mode. This test would require careful checks that the TSH trajectories or TBFs in AIMS do not visit regions of the nuclear configuration not visited by DD-vMCG (and as such not included in the database). Another idea would be to develop a machine-learning version of the cyclobutanone potential energy surfaces (built based on all the trajectories and data collected as part of this challenge evaluated with a consistent electronic-structure method) and use these unified potentials as a way to compare the dynamics obtained with all the mixed quantum/classical and TBF-based methods. Another potentially interesting next step would be to use the best candidate for the electronic-structure calculation and rerun all methods at this very same level. This idea, while in principle trivial, requires great care in its implementation (as highlighted in a recent effort on benchmarking nonadiabatic methods\cite{cigrang2025roadmapNAMD}): initial conditions between methods should be similar, the way to calculate nonadiabatic coupling terms should be unified, the same electronic-structure code should be used, and the interface between the electronic-structure code and the nonadiabatic dynamics code should be the same. Last but not least, the photochemical reactivity of cyclobutanone would constitute an interesting test case to study the impact of zero-point energy leakage in trajectory-based methods for nonadiabatic molecular dynamics.

An interesting theoretical question also emerges from the differences observed between the DD-vMCG results and other nonadiabatic dynamics methods. To undergo ring opening on the timescale seen with other nonadiabatic dynamics methods requires symmetry breaking by excitation of vibrational modes not initially excited in the vertical photoexcitation approach used in the DD-vMCG simulations. This is due either to coupling between vibrations, which was not observed in the DD-vMCG simulations, or some other factor such as the effect of the excitation pulse. Investigation of this issue will be important for future simulations using DD-vMCG.

\subsection{Calculation of observables}
\label{sec:obs}

\begin{figure}[ht]
\centering
\includegraphics[width=0.6\textwidth]{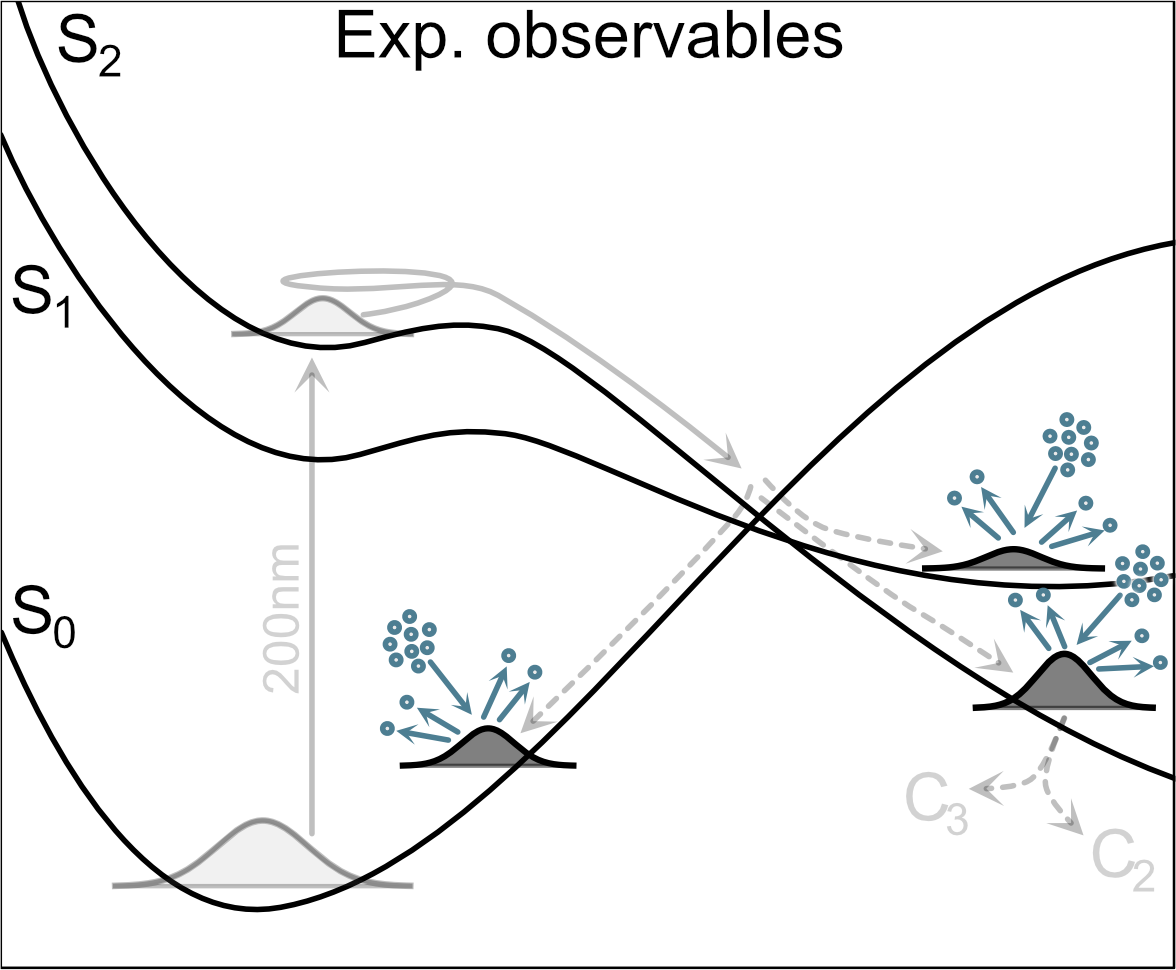}
\caption{Schematic representation of the determination of MeV-UED signal on the support of the nonadiabatic molecular dynamics.}
\label{fig:schemeobs}
\end{figure}

Exp1 sampled the photodynamics of cyclobutanone up to 2 ps following photoexcitation (200 nm, 130 fs FWHM) and obtained the time-resolved difference diffraction signal $\Delta I/I(s,t)$ (with $s$ being the momentum transfer).\cite{green2025cycloUEDExp1} Exp2 studied the first 1.2 ps of dynamics following photoexcitation of cyclobutanone (199.5 nm, 120 fs FWHM) and reported both $\Delta I/I(s,t)$ traces and the time-resolved difference in pair-distribution function $\Delta \text{PDF}(r,t)$ ($r$ is a real-space distance), obtained from the Fourier-sine transform of the difference molecular diffraction intensity $\Delta sM(s,t)$.\cite{wang2025cycloUEDExp2}

The time-resolved experimental signals discussed above can be calculated from the results of nonadiabatic molecular dynamics simulations (see Fig.~\ref{fig:schemeobs} for a schematic depiction). The simplest approach is to use the independent atom model (IAM), where a diffraction signal can be obtained solely based on the position of the nuclei for a given molecular geometry (with no need for any additional electronic-structure information), and all contributions to this challenge followed this route. For trajectory-based approaches, the signal is calculated for each trajectory (which are nothing but time series of molecular geometries) and averaged over the swarm of trajectories, while the signal is obtained from the nuclear wavefunctions for methods using TBFs (within the respective practical approximations of DD-vMCG, AIMS, or AIMC). No diffraction signal was presented from MCTDH simulations. Various strategies were used to extract information from the theoretical colormaps produced for $\Delta I/I(s,t)$ and $\Delta \text{PDF}(r,t)$ -- lineouts, decomposition of the signal per photoproduct, fitting the evolution over time of the extremum of each spectral feature of the $\Delta I/I(s,t)$ and $\Delta \text{PDF}(r,t)$. Paper 3 also calculated a time-resolved photoelectron (TRPES) signal from the AIMS results to compare with the experimental results of Ref.~\citenum{Kuhlman2012a}.

Both experimental $\Delta I/I(s,t)$ signals report a positive and intense feature at very low momentum transfer values $s < 1.2$ \AA$^{-1}$, appearing immediately after photoexcitation and decaying with time constants of 0.29 $\pm$ 0.2 ps (Exp1) and 0.23 ps (Exp2). This feature was attributed to inelastic scattering caused by the $3s$-Rydberg character of the S$_2$ electronic state of cyclobutanone in the Franck--Condon region. This feature was missed by most theoretical contributions due to their use of the IAM, providing only the elastic contribution to the scattering signal. Papers 3 and 8 calculated, for selected geometries, the role of inelastic scattering effects, and highlighted its potentially important contribution for low $s$ values. The description of the inelastic scattering contribution to the $\Delta I/I(s,t)$ signal may be sensitive as well to the precise level of electronic-structure theory employed.\cite{kirrander2019abinitioscatt}

\begin{figure}[ht]
\centering
\includegraphics[width=1.0\textwidth]{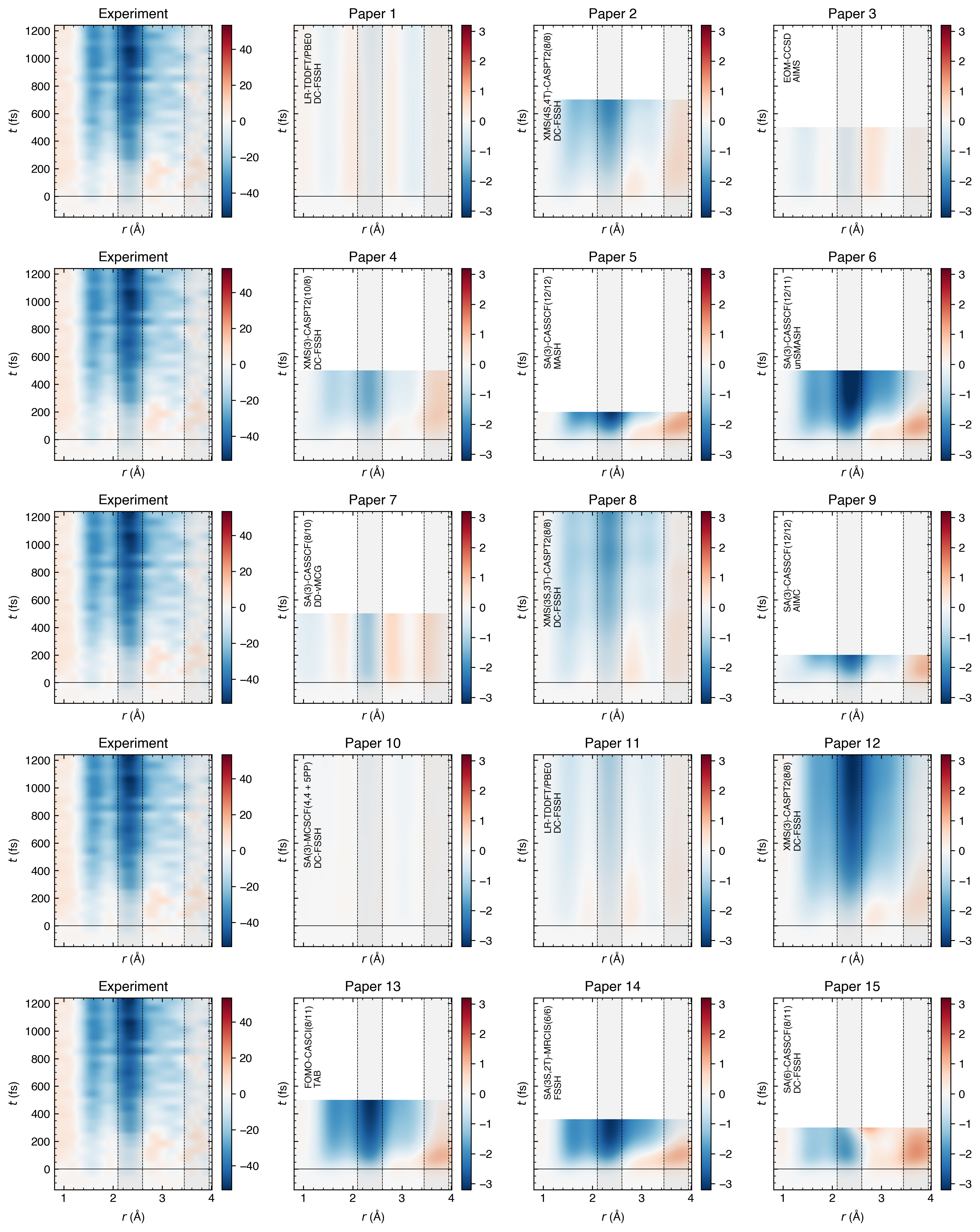}
\caption{Comparison of experimental\cite{wang2025cycloUEDExp2} and predicted UED $\Delta r\text{PDF}(r,t)$ signals. The vertical dashed lines highlight the integration range for the two lineouts plotted in Figs.~\ref{fig:lineouts} and \ref{fig:lineouts2}.}
\label{fig:mosaic}
\end{figure}

As a broad overview of the results of this prediction challenge, we propose here to focus on Fig.~\ref{fig:mosaic} which summarizes the $\Delta r\text{PDF}(r,t)$ predicted by the various contributions and compares them to the results obtained in Exp2. The theoretical signals were all recalculated with the same code (IAM-XED\cite{iamxed}) for the present Perspective, allowing us to compare both the shape and magnitude of the theoretical difference signals (Fig.~\ref{fig:mosaic2} in the Appendix shows another version of Fig.~\ref{fig:mosaic}, with a colormap range optimized for each contribution to improve the visibility of the features, yet preventing a direct comparison between predictions). The difference PDF signal was calculated as $\Delta r\text{PDF}(r,t) = \int_{0}^{10}s\Delta M(s)\sin(sr)\exp(-ks^2)\mathrm{d}s$ with $k=\SI{0.03}{\angstrom^2}$ and the integration range spanning 0 to \SI{10}{\angstrom^{-1}}. The $\Delta M(s)=\Delta I_\mathrm{mol}(s)/I_\mathrm{at}(s)$ function was calculated with the IAM (see the IAM-XED code manual\cite{iamxed} for details). The produced signal was finally convolved using the Gaussian instrument response function with a full width at half maximum parameter of \SI{130}{\femto\second}.  

\begin{figure}[ht]
\centering
\includegraphics[width=1.0\textwidth]{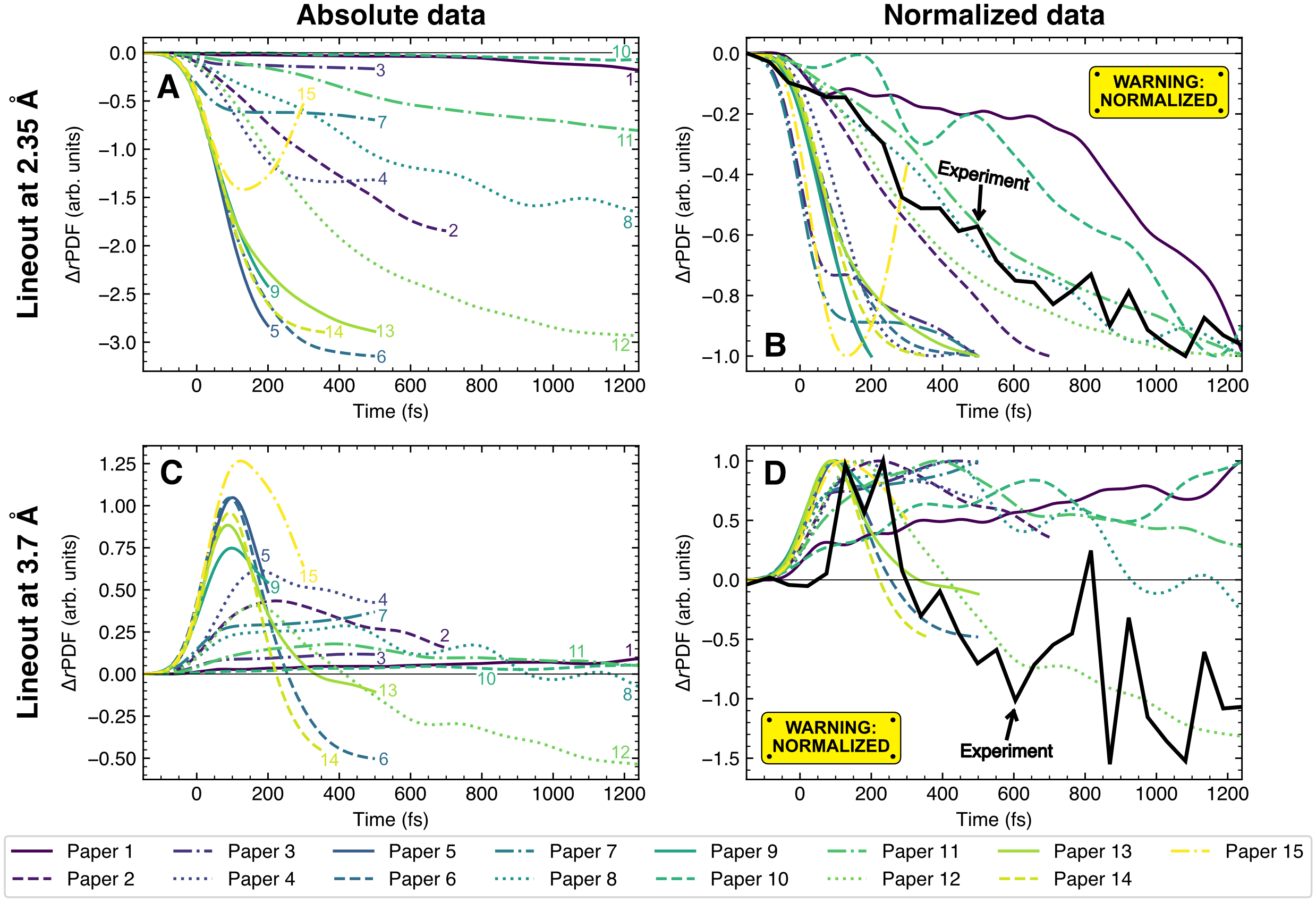}
\caption{Lineouts of the UED signals presented in Fig.~\ref{fig:mosaic}. (A) Absolute values of $\Delta r\text{PDF}$ lineouts at 2.35~Å. (B) Normalized $\Delta r\text{PDF}$ lineout at 2.35~Å. The lineouts were normalized such that the maximum absolute value in the plotted range is equal to one. (C) Absolute values of $\Delta r\text{PDF}$ lineouts at 3.70~Å. (D) Normalized $\Delta r\text{PDF}$ lineout at 3.70~Å. The lineouts were normalized such that the maximum value in the plotted range is equal to one. The experimental lineouts are provided only with the normalized data, as the absolute values of the experimental $\Delta r\text{PDF}$ are not directly comparable with the absolute values of the theoretical predictions.}
\label{fig:lineouts}
\end{figure}

This comparative analysis allows us to stress some of the final findings of this exercise. First of all, most contributions employing an electronic-structure method including static correlation obtained in general a good description of most features of the $\Delta r\text{PDF}(r,t)$, yet with significant differences in the evolution timescales for the main positive and negative features of this observable. XMS-CASPT2-based nonadiabatic molecular dynamics simulations lead, overall, to a better agreement with experiment for the overall trends and lifetime of features appearing on the $\Delta r\text{PDF}(r,t)$ map (e.g., predicted signals in Fig.~\ref{fig:mosaic} for Papers 2, 8, 12). Interestingly, some inconsistency between MCSCF methods can be observed: predictions 7, 10 and 15, do not exhibit the same features as the rest of the predictions employing MCSCF methods (for Paper 15, instabilities in the active space were noted). This observation further highlights that MCSCF methods, namely SA-CASSCF in the present case, are not black box tools and specific settings of parameters, such as active space or number of states, strongly determine the outcome of the calculations. The predictions based on single-reference methods (predicted signals in Fig.~\ref{fig:mosaic} for Papers 1, 3, 11) mostly exhibit a nearly static signal emanating from dynamics near the S$_2$(3s) excited-state minimum, lacking the dissociation features observed with predictions using a method including static correlation. 
Despite the apparent discrepancies between the predicted $\Delta r\text{PDF}(r,t)$ signals represented as colormaps, the reader should bear in mind that the quantitative analysis offered by most contributions (and depicted in Fig.~\ref{fig:lineouts} with lineouts) is already useful to understand the combination of processes occurring during the photodynamics of cyclobutanone.

\begin{figure}[ht]
\centering
\includegraphics[width=1.0\textwidth]{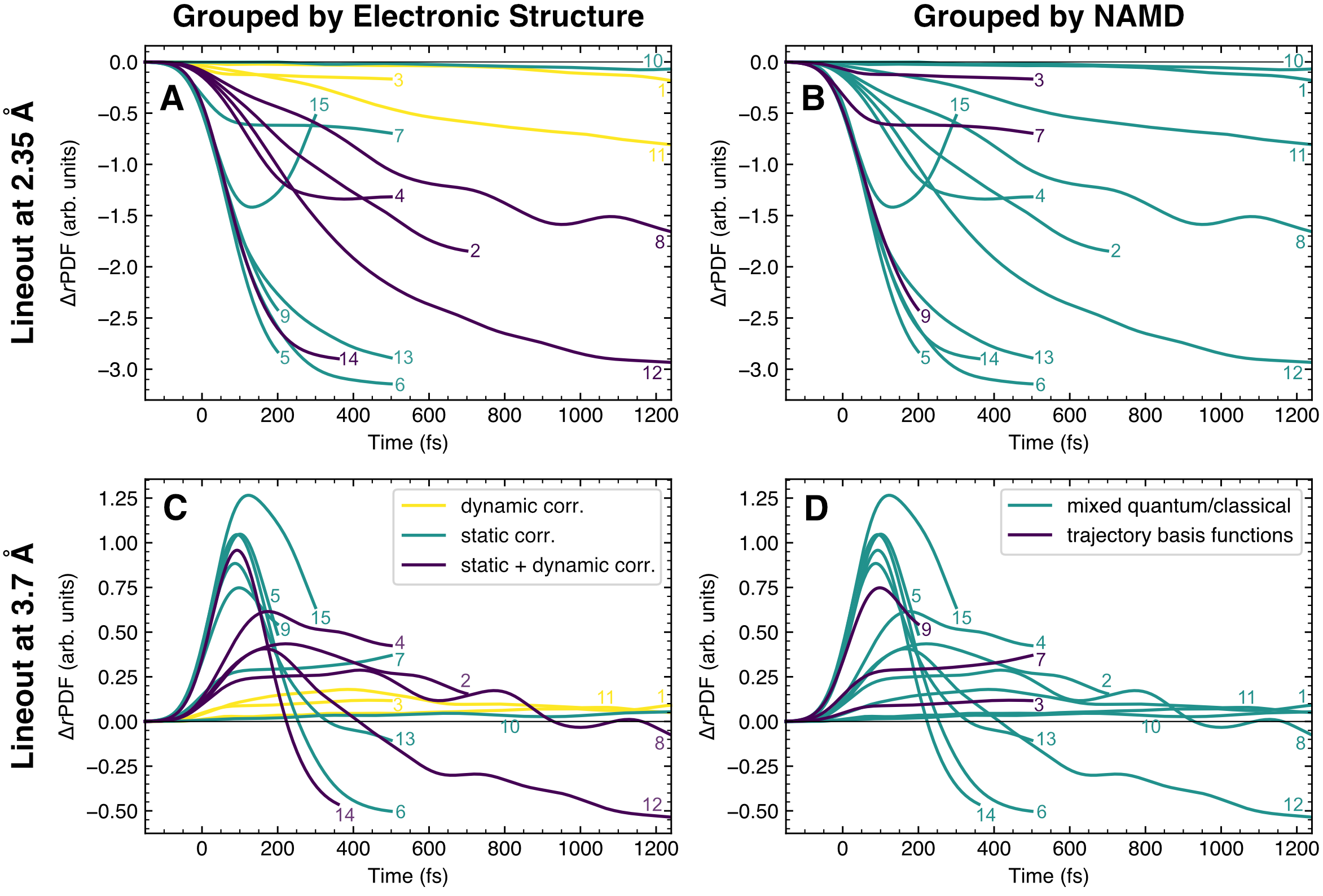}
\caption{Analysis of the lineouts presented in Fig.~\ref{fig:lineouts}A and C, grouping the contributions based on the type of correlation included in the electronic-structure method employed (left column) and on the family of methods used for the nonadiabatic dynamics (right column). (A) Electronic-structure methods, lineout at 2.35~Å. (B) Nonadiabatic dynamics methods, lineout at 2.35~Å. (C) Electronic-structure methods, lineout at 3.70~Å. (D) Nonadiabatic dynamics methods, lineout at 3.70~Å.}
\label{fig:lineouts2}
\end{figure}

While the $\Delta r\text{PDF}(r,t)$ maps plotted in Fig.~\ref{fig:mosaic} are useful to depict the overall time-evolution of the UED signal, they do not permit a clear evaluation of the temporal evolution of their respective features. To unravel the time-dependence of key features in the $\Delta r\text{PDF}(r,t)$ signal, we provide lineouts at \SI{2.35}{\angstrom} and \SI{3.7}{\angstrom} in Fig.~\ref{fig:lineouts}, using both absolute (Figs.~\ref{fig:lineouts}A and C) and normalized (Figs.~\ref{fig:lineouts}B and D) intensities. All lineouts are integrated in a range of \SI{0.5}{\angstrom} to reduce experimental noise and mimic the standard analysis of an experimental signal. The normalized data (normalized with respect to the maximum value along each lineout) are provided here only to facilitate a comparison with the experimental data, but the comparison between absolute scale and normalized data highlights the danger of such a comparison (see end of this paragraph). The lineout at \SI{2.35}{\angstrom} is connected to the ring opening of cyclobutanone, while that at \SI{3.7}{\angstrom} is connected to the fragmentation of the molecule and formation of photoproducts. Fig.~\ref{fig:lineouts}B reinforces the observation of the previous paragraph that predictions employing an electronic-structure method including only static correlation predict too fast dynamics, whereas those making use of an approach combining static and dynamic correlation provide a closer match with the experimental time scale. This dramatic impact of the electronic-structure method on the outcome of the nonadiabatic molecular dynamics is clearly highlighted by Fig.~\ref{fig:lineouts2}A (and Fig.~\ref{fig:lineouts2}C), which represents the same lineouts as in Fig.~\ref{fig:lineouts} but now with each contribution colorcoded based on the type of correlation included in the electronic-structure method employed for the nonadiabatic dynamics (dynamic correlation, static correlation, dynamic+static correlation). Fig.~\ref{fig:lineouts2}A clearly highlights the trends (discussed earlier for the ring-opening process with a few exceptions): electronic-structure methods including solely dynamic correlation (yellow curves) lead to a slower adiabatic ring-opening process in S$_2$ in comparison to nonadiabatic dynamics using approaches including both static and dynamic correlation (purple curves); methods including static correlation only (cyan curves), on the contrary, induce a much faster ring-opening dynamics than that observed with methods including both static and dynamic correlation. The signal obtained in Paper 14 appears as an outlier, which is possibly connected to the fact that MRCIS includes less dynamic correlation than the MRPT approaches. 
However, a correct timescale for a feature of the signal does not mean a correct prediction: Fig.~\ref{fig:lineouts}A shows that the predictions can have significant variations in signal intensities, highlighting an important variation of the overall population of trajectories leading to a given signal. For example, the time trace of the normalized lineout at \SI{2.35}{\angstrom} predicted by Paper 10 using MCSCF appears to have a reasonable timescale in Fig.~\ref{fig:lineouts}B, but this feature is associated with a nearly zero intensity in the absolute signal (Fig.~\ref{fig:lineouts}A). 
The initial increase of the experimental signal at \SI{3.7}{\angstrom} (Fig.~\ref{fig:lineouts}C and Fig.~\ref{fig:lineouts}D) is due to the bond-elongation event during ring opening and dissociations. The following negative signal at later times is caused by the formation of photoproducts, which lack the signal associated with larger bond distances observed in cyclobutanone between carbons and oxygen. This feature is captured only by three predictions, namely Paper 6 (SA-CASSCF), 12 (XMS-CASPT2), and 14 (MRCIS), which simulated the ground-state dynamics of cyclobutanone and its photoproducts for a longer timescale. 
Finally, and keeping in mind the biased distribution towards mixed quantum/classical methods during this challenge, it is interesting to note that the trends observed in the lineouts do not seem to be correlated with the family of methods used for the nonadiabatic molecular dynamics (Fig.~\ref{fig:lineouts2}B and Fig.~\ref{fig:lineouts2}D). Additional simulations would be required to further support this lack of correlation.

\subsubsection{To-do list: Calculation of observables}
\label{sec:obstd}

The most pressing follow-up on these calculations is the careful incorporation of the inelastic scattering contribution to the electron scattering signal. Two contributions proposed initial calculations, but given the complexity of this property -- and its potential sensitivity to the level of electronic-structure theory -- a detailed and careful benchmark study is needed. 
More direct comparisons between theory and experiment will be required, possibly with the development of new strategies to compare one-to-one predicted and experimental (time-resolved) observables as information-rich as scattering data. One type of experiment is also not enough to ensure that the photodynamics of cyclobutanone is fully characterized and to assess more broadly the quality of the nonadiabatic dynamics simulations. Other experimental observables related to time-resolved valence photoelectron spectroscopy or X-ray absorption/photoelectron spectroscopy could provide more information about the electronic states involved in the nonradiative decay of cyclobutanone and unravel further differences between the different nonadiabatic dynamics simulations reported here, invisible from the Perspective of a scattering experiment.

\section{Present understanding of the photochemistry of cyclobutanone}

\begin{figure}[ht]
\centering
\includegraphics[width=0.6\textwidth]{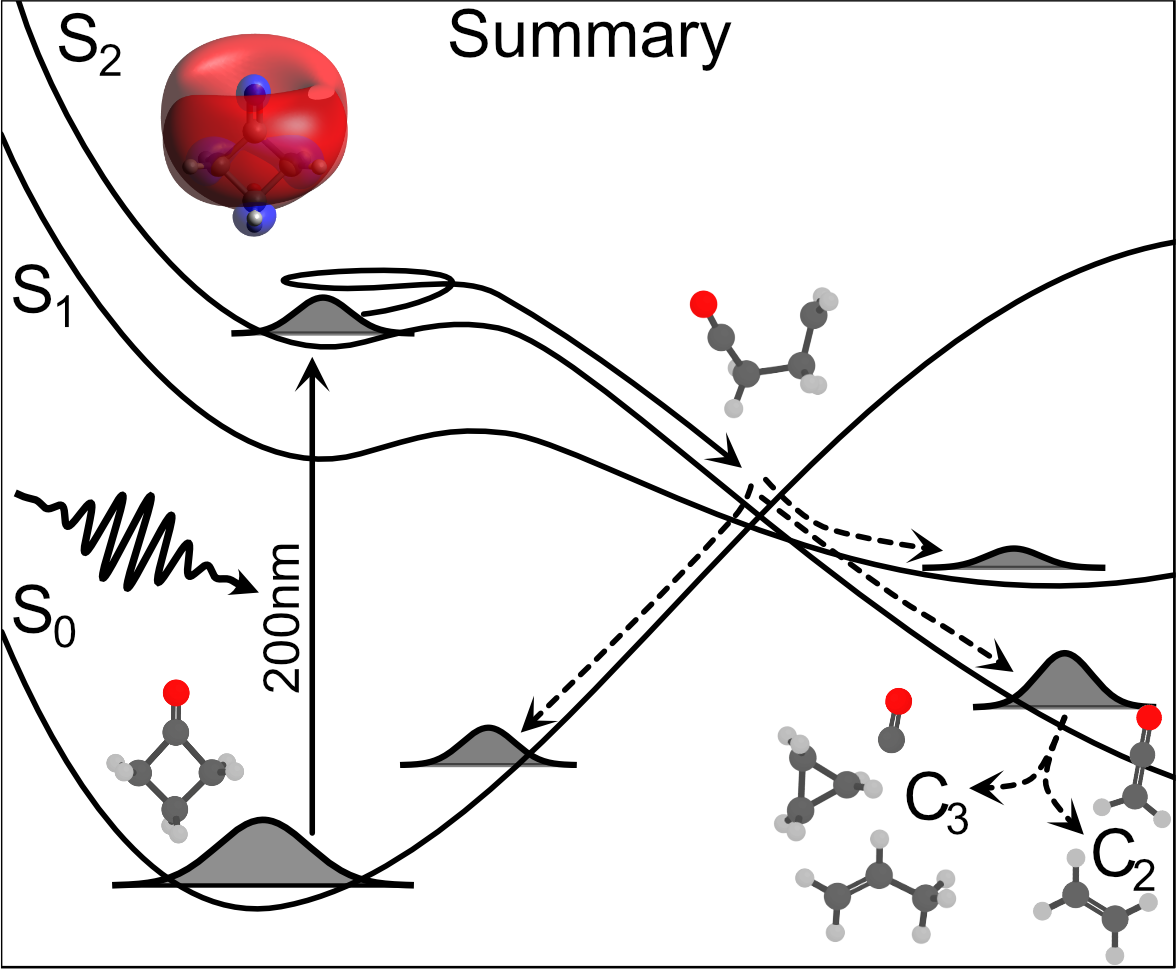}
\caption{Schematic representation of the overall photochemistry of cyclobutanone excited at 200 nm.}
\label{fig:schemesum}
\end{figure}

Based on the various contributions to this challenge, the discussion between contributors during the dedicated CECAM workshop, and the comparisons with the MeV-UED experimental results, the photochemistry of cyclobutanone excited at 200 nm with a laser pulse takes place as follows. The pump laser pulse photoexcites cyclobutanone to its second excited electronic state S$_2$, exhibiting $3s$-Rydberg character in the Franck--Condon region. The nuclear wavepacket formed in this electronic state remains in the vicinity of the Franck--Condon region for a few hundred femtoseconds. This first step, photophysical in nature, contributes to the vibronic progression observed in the UV/Vis photoabsorption cross-section of the molecule, as determined by contributions using MCTDH. Photoexcited cyclobutanone then undergoes photochemical processes, evolving \textit{adiabatically} in S$_2$ from a Rydberg to a valence state, requiring the passage across a barrier in S$_2$ and ring opening via the carbon-carbon bond near the carbonyl group, driving the excited molecule towards S$_1$ and S$_0$. Part of the nuclear wavepacket can also reach an S$_2$/S$_1$ conical intersection (or, more precisely, its intersection seam) where the molecule remains ring-closed and only subsequently ring-opens in S$_1$. Nonadiabatic transitions to the ground electronic states then occur, leading to the formation of the various (major) photoproducts: CO and cyclopropane, CO and propene, ethene and ketene, cyclobutanone. No contribution found evidence that intersystem crossing plays a major role on the sub-ps to ps timescales probed here for the photochemistry of cyclobutanone when excited at 200 nm (photoexcitation of cyclobutanone into its S$_1(n\pi^\ast)$ state is likely to change this picture dramatically as observed in earlier solution-phase studies,\cite{kao2020effects} but was not studied here). These findings address most of the 'question marks' in Fig.~\ref{fig:schemecyclo} -- the original scheme depicting the potential photochemistry of cyclobutanone -- and are depicted schematically in Fig.~\ref{fig:schemesum}.

\section{Summary of the results and Outlook}
\label{sec:summary}

The precise computational details for the various topics discussed above impacted, in the first place, the time scales for the various processes constituting the photochemical mechanism. The proportion of photoproducts (photoproduct yields) is perhaps harder to decipher given that these quantities were determined at different times during the dynamics (from a few tens of fs to 2 ps following photoexcitation), but also seem to be influenced by the level of electronic-structure theory, yet to a lesser extent. A clear omission of all the predictions submitted is a UED feature that was assigned experimentally to inelastic scattering from the S$_2(3s)$ electronic state at low $s$, and stresses the extra challenge of calculating accurate experimental observables on top of already-complex nonadiabatic molecular dynamics simulations. 

During the CECAM workshop, it was proposed that this challenge can be seen as a \textit{calibration} exercise. The nature of nonadiabatic dynamics simulations is such that the user is left with a wide range of parameters and approaches to choose from, as is made explicit by the complexity of tables summarizing the results. This challenge offered an opportunity to explore the enormous parameter space of a nonadiabatic molecular dynamics simulation and, for the strategies deployed here, to define a set of practical guidelines and caveats. Perhaps the most important (and somewhat expected) outcome of this challenge is the critical influence of the level of electronic-structure theory, and the need to benchmark it in and beyond the Franck--Condon region. Unfortunately, the precise impact of the method used for the nonadiabatic dynamics in this Prediction Challenge is, and probably will remain, difficult to assess. As suggested above, the ideal test is to rerun all methods employing the best candidate for the electronic-structure calculation, an effort that might require developments beyond the scope of the challenge as well as a computational cost that is too large for some methodologies. This prediction exercise also stimulated discussions about reproducibility and comparability in computational photochemistry.\cite{list20226pce}

This challenge also highlights the potential influence that the description of the photoexcitation process can have on the nonadiabatic dynamics, and the sensitivity of the experimental observable to the underlying excited-state dynamics. This Perspective summarizes additional guidance related to reproducibility in the field of nonadiabatic molecular dynamics. We can expect that a significant number of studies will emerge on cyclobutanone as a follow-up to this challenge (we note a recent work combining TSH and MR-SF-TDDFT to study the photodynamics of cyclobutanone,\cite{brady2025mrsf} including higher Rydberg states). Finally, this calibration exercise also highlights the importance of working as a community to identify the key factors influencing the outcome of a nonadiabatic molecular dynamics. A single study on the photochemistry of cyclobutanone would most likely not have been able, on its own, to converge to a reasonable result and form an educated guess -- a prediction -- about the photochemistry of cyclobutanone with great certainty. Combining the outcome of 15 contributions helped decipher the important parameters for this type of simulations and was necessary, at least as a calibration exercise, to draw the general recommendations presented in this Perspective and to support future predictive endeavors.

The community may also want to invest more effort on assessing the level of uncertainty in nonadiabatic molecular dynamics, supporting the overarching goal of obtaining decent predictions at a reasonable computational cost. It is interesting to note that most contributions to this challenge included a paragraph assessing the validity and limitations of the methodologies used to produce their final prediction. This kind of paragraph could become a new standard in articles reporting nonadiabatic molecular dynamics simulations, allowing the reader to assess the reliability of the results presented.

This Prediction Challenge also paves the way for future community activities. A central action is to interact further with developers of electronic-structure methods and discuss some of the challenges observed for this simple-looking molecule, in particular beyond the Franck--Condon region -- CECAM workshops and discussions, possibly with the Psi-k network, would offer an optimal venue to trigger such interactions and stimulate long-term collaborations between (nonadiabatic) dynamicists and developers of electronic-structure methods. Given the calibration nature of this first exercise, the contributors suggested the organization of a second challenge with a more controlled set of parameters. A possible idea would be to solely test the methods for nonadiabatic dynamics on model systems (with fixed and common electronic-structure quantities for all simulations), for which quantum dynamics results are reachable. This project could connect to the benchmarking effort in nonadiabatic dynamics (another event supported by CECAM) summarized in Ref.~\citenum{cigrang2025roadmapNAMD}. Another idea for a future challenge would be to develop a coordinated experimental campaign designed to probe and challenge specific aspect of the theoretical methodologies associated to nonadiabatic molecular dynamics -- in other words, using experimental facilities as dedicated stress tests for theory.
Efforts to standardize the interfaces between nonadiabatic dynamics methods and electronic-structure packages are also greatly needed, and in progress with codes like Newton-X\cite{barbatti2022newtonx} and the development of the COSMOS project\cite{cosmosproject} around Quantics\cite{worth2020quantics}. From a community perspective, the time pressure associated with this challenge may not have allowed all interested groups to contribute. (Unfortunately, not all contributors received the email about a one-month deadline extension.) 
Ideally, future challenges would have more advance notice, facilitating broader participation within the community.  

Finally, coming back to the original question that motivated this challenge -- Is the field of nonadiabatic molecular dynamics mature enough to be predictive? -- one may be tempted to answer \textit{nearly}. This exercise showed that nonadiabatic molecular dynamics can be qualitatively predictive: Fig.~\ref{fig:mosaic} showed that many predicted UED signals reproduced the main features of the experimental UED time trace and would provide valuable insight to guide the analysis of the experimental signal. Nearly quantitative reproduction of major experimental UED features was achieved with nonadiabatic dynamics using electronic-structure methods including both static and dynamic correlation. From a chemical perspective, most contributions using an electronic-structure approach including static correlation obtained the same families of photoproducts, mostly formed on the ground electronic state following nonradiative decay. All contributions agreed that intersystem crossing processes are not important for the photochemistry of gas-phase cyclobutanone photoexcited at 200 nm. The discussions contained in this Perspective also provide important guidelines for benchmarking that should support the predictive power of future nonadiabatic dynamics simulations of gas-phase molecules (such guidelines are also further detailed in recent works\cite{bestpractices2026}). 
Nevertheless, we contrast this positive outcome with more realistic considerations. First, this challenge focused on a single experimental observable from MeV-UED measurements, which may have been more forgiving to some methods than other experimental observables would have been. Second, this challenge encompassed both photophysical and photochemical processes, which constitute a stringent test for nonadiabatic molecular dynamics. However, these processes took place in the gas phase --advances in liquid jets at advanced light sources mean that the nonadiabatic dynamics community may soon need to consider similar tests for ultrafast dynamics in complex environments.

\begin{acknowledgments}
The authors would like to thank CECAM for supporting the organization of the CECAM workshop 'Triggering out-of-equilibrium dynamics in molecular systems' in March 2023, which led to the idea of this prediction challenge, as well as the CECAM workshop 'Building a roadmap for future developments in computational photochemistry based on a prediction challenge' that took place in April 2025 and resulted in this article. The authors also thank Prof. Dao Xiang for providing the experimental MeV signal used for the comparison between theory and experiment presented in this work. The authors finally acknowledge the Virtual Winter School on Computational Chemistry for fruitful discussions connected to the results of this challenge.
BFEC acknowledges funding from the European Research Council (ERC) under the European Union's Horizon 2020 research and innovation programme (Grant agreement No. 803718, project SINDAM). 
BFEC, DH, GAW, AK, DS, JE, KES, DVM, and TJP acknowledge the COSMOS EPSRC Programme Grant (EP/X026973/1).
BFEC, GAW, and AJOE acknowledge the UPDICE EPSRC Programme Grant (EP/V026690/1).
BFEC and JCG acknowledge the EPSRC Grant EP/Y01930X/1.
JJ and PS were supported by the Czech Science Foundation, project number 23-07066S. This work was supported from the grant of Specific university research – grant No A2\_FCHI\_2025\_053. 
JE and TJP made use of the Rocket High Performance Computing service at Newcastle University. TJP would like to thank the EPSRC for an Open Fellowship (Grant No. EP/W008009/1).
OJF is supported by the U.S. Department of Energy, Office of Science, Office of Advanced Scientific Computing Research, Department of Energy Computational Science Graduate Fellowship under Award Number DE-SC0023112. OJF and TJM are supported by the AMOS program of the U.S. Department of Energy, Office of Science, Basic Energy Sciences, Chemical Sciences, Geosciences, and Biosciences Division.
NHL acknowledges funding from the Swedish Research Council (Grant no.~2022-02871).
MG acknowledges support from the AMOS program of the U.S. Department of Energy, Office of Science, Office of Basic Energy Sciences, Division of Chemical Sciences, Geosciences, and Biosciences under Award Number DE-SC0022225.
LH acknowledges support from the Leverhulme Trust (Grant No. RPG-2020-208).
FA acknowledges funding from the French Agence Nationale de la Recherche via the projects Q-DeLight (Grant No. ANR-20-CE29-0014) and STROM (Grant No. ANR-23-ERCC-0002).
JS and BGL gratefully acknowledge support from the National Science Foundation under Grant No. CHE-1954519 and the Institute for Advanced Computational Science (IACS).
AEG thanks the European Union through Horizon Europe Project No. 123-CO:101067645.
FJH acknowledges the support from the School of Physical and Chemical Sciences at Queen Mary University of London and the support from the QMUL Research-IT.
XM thanks Studienstiftung des deutschen Volkes for a fellowship.
PVZ acknowledges support from the NSF (CHE-2054616 and CHE-2054604), the Simons Foundation (computational resources), New York University (RC110 Postdoctoral Research and Professional Development Support Grant), and NYU IT High Performance Computing.
ZL acknowledges support from the National Natural Science Foundation of China (No. 22333003 and 22361132528).
SG thanks the COST action CA21101 ``COS'' supported by COST (European Cooperation in Science and Technology) and the Spanish Ministry of Science and Innovation under grant PID2024-155352NB-C21.
MB acknowledges the European Research Council (ERC) Advanced grant SubNano (Grant Agreement
No. 832237).
MBr and RCO acknowledge funding from the UCL DTP (EP/W524335/1).
JEL was supported by an Independent Postdoctoral Fellowship at the Simons Center for Computational Physical Chemistry, under a grant from the Simons Foundation (839534, MT).
JRM acknowledges support from the Alexander von Humboldt Foundation.

\end{acknowledgments}

\clearpage
\begin{appendix}
\section{UED signal plotted with optimized colormap for each prediction}

\begin{figure}[ht]
\centering
\includegraphics[width=0.825\textwidth]{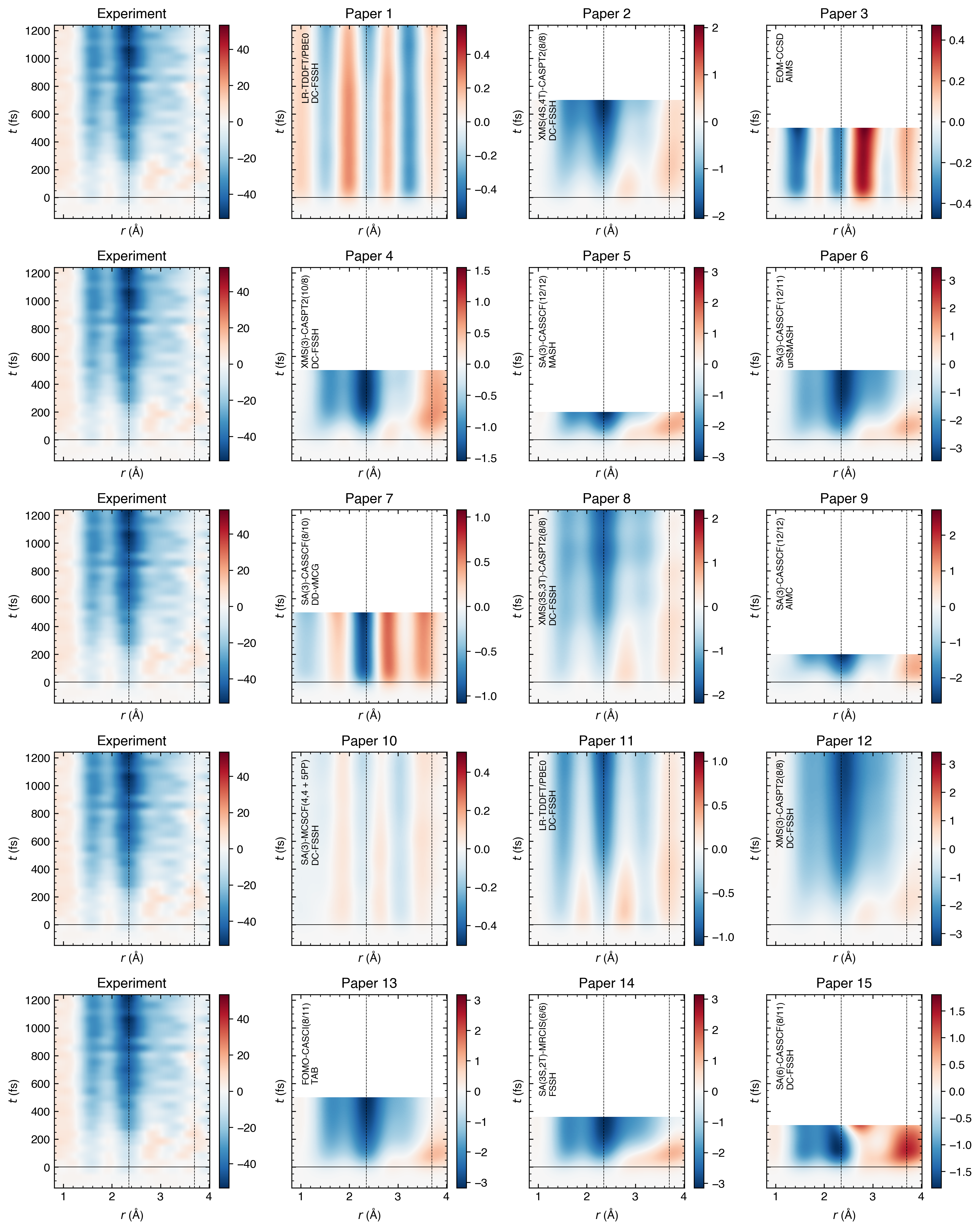}
\caption{Comparison of experimental\cite{wang2025cycloUEDExp2} and predicted UED $\Delta r\text{PDF}(r,t)$ signals. The colormap range was optimized for each contribution to improve the visibility of the different features. As a result, the color intensities are not comparable one-to-one between the theoretical predictions (in contrast to Fig.~\ref{fig:mosaic}).}
\label{fig:mosaic2}
\end{figure}

\end{appendix}

\section*{Data Availability Statement}

The data that support the findings of this study are openly available in Zenodo at \url{https://doi.org/10.5281/zenodo.20559411}


%

\end{document}